\documentclass[journal]{IEEEtran}
\usepackage[utf8]{inputenc}
\usepackage{cite}
\usepackage{graphicx}
\usepackage{amsmath}
\usepackage{mathrsfs,amsthm,amsmath,amssymb}
\usepackage{textcomp}
\usepackage{graphicx,multirow}
\usepackage{balance}
\usepackage{float}
\usepackage{gensymb}
\usepackage{eso-pic}
\usepackage{algorithm}
\usepackage{algorithmic}
\usepackage{academicons}
\usepackage{xcolor}
\usepackage{ulem}
\usepackage{comment}
\usepackage{orcidlink}
\usepackage{caption}
\usepackage{booktabs}
\makeatletter
\newcommand\subparagraph{%
  \@startsection{subparagraph}{5}
  {\parindent}
  {3.25ex \@plus 1ex \@minus .2ex}
  {-1em}
  {\normalfont\normalsize\bfseries}}
\makeatother
\usepackage{titlesec}
\let\subparagraph\relax

\usepackage{titlesec}
\usepackage{scalerel}
\usepackage{tikz}
\usepackage{subcaption}

\usepackage[english]{babel}

\newcommand{\mb}{\mathbf}

\usepackage{bm} 

\DeclareMathOperator{\diag}{diag}

\usetikzlibrary{svg.path}

\definecolor{orcidlogocol}{HTML}{A6CE39}
\tikzset{
  orcidlogo/.pic={
    \fill[orcidlogocol] svg{M256,128c0,70.7-57.3,128-128,128C57.3,256,0,198.7,0,128C0,57.3,57.3,0,128,0C198.7,0,256,57.3,256,128z};
    \fill[white] svg{M86.3,186.2H70.9V79.1h15.4v48.4V186.2z}
                 svg{M108.9,79.1h41.6c39.6,0,57,28.3,57,53.6c0,27.5-21.5,53.6-56.8,53.6h-41.8V79.1z M124.3,172.4h24.5c34.9,0,42.9-26.5,42.9-39.7c0-21.5-13.7-39.7-43.7-39.7h-23.7V172.4z}
                 svg{M88.7,56.8c0,5.5-4.5,10.1-10.1,10.1c-5.6,0-10.1-4.6-10.1-10.1c0-5.6,4.5-10.1,10.1-10.1C84.2,46.7,88.7,51.3,88.7,56.8z};
  }
}

\newcommand\orcidicon[1]{\href{https://orcid.org/#1}{\mbox{\scalerel*{
\begin{tikzpicture}[yscale=-1,transform shape]
\pic{orcidlogo};
\end{tikzpicture}
}{|}}}}

\graphicspath{ {Images/} }
\IEEEoverridecommandlockouts

\begin{document}
\setlength{\parskip}{5pt}
\setlength{\abovedisplayskip}{5pt}
\setlength{\belowdisplayskip}{5pt}

\title{Direct Uplink Connectivity in Space MIMO Systems with THz and FSO Inter-Satellite Links}

\author{Zohre~Mashayekh~Bakhsh$^\text{\orcidlink{0000-0002-3798-968X}}$,
        Yasaman~Omid$^\text{\orcidlink{0000-0002-5739-8617}}$,
        Gaojie~Chen$^\text{\orcidlink{0000-0003-2978-0365}}$,~\IEEEmembership{Senior Member,~IEEE,}
        Farbod~Kayhan$^\text{\orcidlink{0000-0002-1733-5435}}$,
        Yi~Ma$^\text{\orcidlink{0000-0002-6715-4309}}$,~\IEEEmembership{Senior Member,~IEEE,}
        and Rahim~Tafazolli$^\text{\orcidlink{0000-0002-6062-8639}}$,~\IEEEmembership{Senior Member,~IEEE}
 \thanks{Zohre Mashayekh Bakhsh, Gaojie Chen, Yi Ma, and Rahim Tafazolli are with the Institute for Communication Systems, Home of
the 6G Innovation Centre, University of Surrey, GU2 7XH Guildford, U.K. 
Yasaman Omid is with Loughborough University, and Farbod Kayhan is the Lead SatCom 5G R\&D Engineer at Telespazio, UK.
(E-mails: z.mashayekhbakhsh@surrey.ac.uk, y.omid@lboro.ac.uk, gaojie.chen@ieee.org, farbod.kayhan@telespazio.com, y.ma@surrey.ac.uk, r.tafazolli@surrey.ac.uk) (\it Corresponding author: Gaojie Chen.)}%

}

\maketitle
\begin{abstract}
This paper investigates uplink transmission from a single-antenna mobile phone to a cluster of satellites, emphasizing the role of inter-satellite links (ISLs) in facilitating cooperative signal detection.
The study focuses on non-ideal ISLs, examining both terahertz (THz) and free-space optical (FSO) ISLs concerning their ergodic capacity. \textcolor{black}{
We present a practical scenario derived from the recent 3GPP standard, specifying the frequency band, bandwidth, user and satellite antenna gains, power levels, and channel characteristics in alignment with the latest 3GPP for non-terrestrial networks (NTN). Additionally, we propose a satellite selection method to identify the optimal satellite as the master node (MN), responsible for signal processing. This method takes into account both the user-satellite link and ISL channels.}
For the THz ISL analysis, we derive a closed-form approximation for ergodic capacity under two scenarios: one with instantaneous channel state information (CSI) and another with only statistical CSI shared between satellites. For the FSO ISL analysis, we present a closed-form approximate upper bound for ergodic capacity, accounting for the impact of pointing error loss.
Furthermore, we evaluate the effects of different ISL frequencies and pointing errors on spectral efficiency. Simulation results demonstrate that multi-satellite multiple-input multiple-output (MIMO) satellite communication (SatCom) significantly outperforms single-satellite SatCom in terms of spectral efficiency. Additionally, our approximated upper bound for ergodic capacity closely aligns with results obtained from Monte Carlo simulations.

\end{abstract}

\begin {IEEEkeywords}
Direct Device-Satellite Communications,  Multi-Satellite Cooperative Communication, Ergodic Capacity, FSO ISL, THz ISL.   
\end{IEEEkeywords}

\section{Introduction}
\label{intro}
\IEEEPARstart{T}{he} sixth-generation (6G) communication system is expected to support a wide range of services, including higher data rates and global coverage. However, underserved regions, such as rural and developing areas, continue to experience poor internet connectivity. Satellite communications (SatComs), with their extensive coverage capabilities, are well-suited to serve both these rural areas and the existing congested regions \cite{10179219, 9852737}.

High-throughput direct connections between satellites and unmodified handheld devices, such as cell phones, represent a promising technology for achieving global connectivity.
This approach has the potential to provide seamless connectivity worldwide \cite{10646360}.
One possible solution to enhance data transfer rates is through cooperative satellite operations. This strategy could reduce the need for larger antennas and improve data throughput within the same constellation. By fostering collaboration between satellites, the advantages of multiple-input multiple-output (MIMO) technology in SatCom can effectively complement traditional MIMO techniques, thereby boosting overall system performance \cite{9351765}.

For effective collaboration, inter-satellite links (ISLs) are essential as they enable seamless data and channel state information (CSI) sharing between satellites.
ISLs come in various forms, including radio frequency (RF), terahertz (THz), and free-space optical (FSO), each tailored to meet the specific requirements of the communication network.
RF/THz technology benefits from mature development does not require stringent acquisition and tracking functionalities, but is prone to interference and generally offers lower data rates compared to optical media. Conversely, FSO communication links provide higher data rates and feature smaller equipment sizes and lower power requirements. However, they necessitate more complex acquisition and tracking functionalities to maintain effective communication \cite{9852737}.
The physical topological parameters of ISLs, such as distance and pointing direction, change periodically with satellite movement. In the study by \cite{chen2021modeling}, the authors developed a calculation model for these ISL geometrical parameters, focusing on range and pointing direction, which are essential for communication performance. In \cite{maamar2016study}, the authors investigated the link budget and communication performance of optical ISLs.
The authors of \cite{nie2021channel} explored the characteristics of the propagation channel in the THz spectrum.
Moreover, the capacity of THz inter-satellite communication systems utilizing high-gain reflective antennas was analyzed by the authors of \cite{ding2016analysis}.

\begin{table*}[!h]
\centering
\caption{\textcolor{black}{Highlighting the unique aspects of our work in comparison to the relevant literature}}
\label{tab:comparison}
\begin{tabular}{|l|c|c|c|c|c|}
\hline
\textbf{} & \textbf{\cite{Salem2024}} & \textbf{\cite{10197164}} & \textbf{\cite{10279447}} & \textbf{\cite{chen2024direct}} & \textbf{Our Work} \\ \hline
\textcolor{black}{Considering non-ideal ISLs (FSO/THz)}              & $\checkmark$     & $\times$     & $\times$     & $\times$     & $\checkmark$ \\ \hline
\textcolor{black}{Ergodic capacity evaluation considering both THz and FSO ISLs}          & $\times$     & $\times$     & $\checkmark$     & $\times$     & $\checkmark$ \\ \hline
\textcolor{black}{Detection vector design based on both instantaneous CSI and statistical CSI} & $\times$ & $\times$ & $\times$ & $\times$ & $\checkmark$ \\ \hline
\textcolor{black}{Master node satellite selection}                & $\times$     & $\times$     & $\times$     & $\times$     & $\checkmark$ \\ \hline
\textcolor{black}{Practical scenarios (powers, gains, channel model) Based on \cite{3GPPrel15}}     & $\checkmark$     & $\checkmark$     & $\times$     & $\checkmark$     & $\checkmark$ \\ \hline
\end{tabular}
\end{table*}

\section{Related Works}\label{related work}
In this section, we review studies on multi-satellite MIMO systems, encompassing both downlink and uplink scenarios.
\textcolor{black}{The authors of \cite{roper2022beamspace} proposed a position-based beamforming scheme that operates without instantaneous CSI, designing a distributed linear precoding method based solely on the relative position information between multiple satellites and a single user terminal. This approach enables joint transmission with minimal information exchange. The study also identified the optimal inter-satellite distance to maximize data rates. Similarly, the authors of \cite{9814655} extended this model to multiple users and employed the Deterministic Equivalence (DE) method to derive optimal precoding vectors based on statistical CSI, facilitating satellite cooperation to reduce interference.
In \cite{li2022performance}, each satellite serves distinct cells. The authors derived the received signal's probability density function (PDF) and a closed-form expression for multiple access interference (MAI) under different fading models. They observed that the transmission rate increases linearly with a small number of users but slows down as the user count grows. Therefore, a balance between the transmission rate and the number of users is essential for achieving high energy efficiency.}

In \cite{abdelsadek2022distributed}, satellites collaboratively transmit symbols to users using coherent beamforming techniques to reduce inter-user interference.
The authors demonstrated that the multi-satellite approach significantly improves both the average user service time and spectral efficiency compared to using a single satellite.
In \cite{10061620}, the authors expanded on their earlier work from \cite{abdelsadek2022distributed}, analyzing spectral efficiency when both satellites and user terminals employed multiple antennas. The study included a detailed comparison of three scenarios: distributed massive MIMO (DM-MIMO) connectivity, where multiple satellites work together; collocated massive MIMO connectivity, where all antennas are on a single satellite; and single satellite connectivity, involving only one satellite. Simulation results confirmed that the DM-MIMO approach outperforms the other approaches in terms of spectral efficiency.

We focused on the uplink direction in \cite{omid2023oncapacity}, which presents greater challenges in direct user-satellite communications.
Building on the works in \cite{maaref2007joint} and \cite{ratnarajah2004spatially}, we derived the ergodic capacity for an uplink scenario where satellite clusters operate without real-time CSI. Instead, these clusters use shared statistical CSI to implement a joint detection strategy that minimizes mean square error (MSE).
In \cite{omid2023spacemimo}, we explored two scenarios: one where full CSI is shared among satellites and another where only partial CSI is shared. We conducted a comparative analysis of these scenarios in terms of bit error rate (BER), overhead, and capacity.
However, none of the aforementioned studies examined the impact of ISL noise and pointing error loss on system performance. Additionally, different types of ISLs had not been compared in terms of capacity.

\textcolor{black}{The study in \cite{Salem2024} examines the block error rate (BLER) in a handover scenario for uplink transmission from a mobile phone to multiple satellites, considering both FSO and THz ISLs but does not focus on ergodic capacity calculations. In \cite{10197164}, the authors optimize the energy efficiency of uplink transmission from a lens antenna array to multiple satellites, assuming ideal ISLs. The work in \cite{10279447} investigates an uplink transmission from a very small aperture terminal (VSAT) to multiple satellites, designing a precoder to maximize the signal-to-leakage-plus-noise ratio (SLNR) while also assuming ideal ISLs. In \cite{chen2024direct}, uplink transmission from a mobile phone to multiple satellites is analyzed, focusing on the design and evaluation of an orthogonal frequency division multiplexing (OFDM)-based multi-dimensional constellation transmission scheme without accounting for ISL impairments. Moreover, none of these studies consider a satellite selection scheme.
Our contributions are compared to the most relevant existing uplink studies in the literature, as summarized in Table \ref{tab:comparison}, with further details on our innovations provided below.}
\begin{itemize}
    \item \textcolor{black}{\textbf{Practical Space MIMO System Scenario:} We propose a practical Space MIMO system for direct transmission from a single-antenna mobile phone to a cluster of satellites that collaborate to detect the signal. The design aligns with 3GPP standards. 
    Additionally, we adopt realistic user-satellite channel models, including the random Rician \( K \)-factor and shadow fading, based on 3GPP guidelines \cite{3GPPrel15}.}
    \item \textcolor{black}{\textbf{Master Node (MN) Satellite Selection:} To optimize system performance, we propose a method for selecting the best satellite as the MN based on both user-satellite link quality and ISL conditions. Our results confirm the effectiveness of this selection method in enhancing system performance.}
    \item \textcolor{black}{\textbf{Ergodic Capacity in THz/FSO ISL:} We derive closed-form approximations for the ergodic capacity under two scenarios: instantaneous CSI sharing and statistical CSI sharing between satellites when considering the design of optimal detection vector. For FSO, we model the impact of pointing error loss and derive a closed-form approximation for the upper bound of the ergodic capacity.}
\end{itemize}

The structure of this paper is as follows:
Section \ref{System Model} describes the system model and user-satellite link channel.
Section \ref{THz ISL} covers the ergodic capacity analysis with THz ISLs, including instantaneous CSI and statistical CSI scenarios, the ISL channel model, and MN satellite selection.
Section \ref{FSO ISL} examines the ergodic capacity of the Space MIMO system with FSO ISLs, accounting for pointing error losses.
Section \ref{Numerical Results} presents numerical results, and Section \ref{conclusion} concludes the work.

\section{System Model}
\label{System Model}

\begin{figure}[t]
    \centering
    \includegraphics[trim={41cm 0cm 40cm 0cm}, scale=0.6]{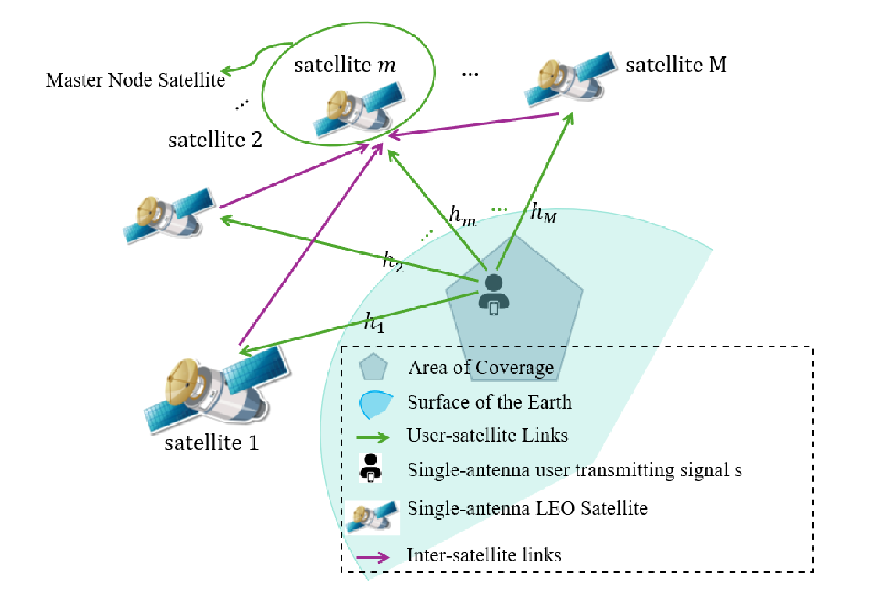}
    \caption{Space MIMO system model.}
    \label{fig:system model}
\end{figure}

This study analyzes uplink transmission from a single-antenna mobile phone to multiple transparent LEO satellites \textcolor{black}{in a suburban environment within the S-band frequency.} These satellites operate at two different altitudes and elevation angles, resulting in variations in the distance between the user and each satellite over time.
At any given time step, the maximum number of satellites within a $30^{\degree}$ elevation angle from the user is denoted as $M_{max}$. The value of $M_{max}$ depends on the satellite constellation and changes dynamically over time. Each satellite communicates with the mobile phone via both line-of-sight (LoS) and non-line-of-sight (NLoS) links.
To enhance system performance and increase capacity, a cluster of satellites is formed using ISLs, which collaboratively process signals received from the mobile phone. The system model is illustrated in Fig. \ref{fig:system model}.

The mobile phone broadcasts its signal to all satellites in the constellation simultaneously. These satellites are ranked by their proximity to the user, with the closest one designated as Satellite 1. Among the visible satellites, one is assigned as the MN satellite, which is responsible for processing the received signals. This MN satellite is regenerative, meaning it can process the data internally.
Assuming the satellites in the constellation are aware of each other's positions through ephemeris data, inter-satellite communication, or ground updates, each satellite can identify the MN satellite to which it should forward its signal. Once the MN satellite is determined, it identifies the 
$M-1$ nearest satellites to itself and collects signals from them. This approach reduces the overall system overhead, allowing the remaining satellites in the constellation to be reallocated to serve other areas on Earth.
The identified relaying satellites transmit their signals to the MN satellite using the amplify-and-forward (AF) relaying technique. AF is preferred over decode-and-forward (DF) due to its lower complexity for relaying satellites. Moreover, as demonstrated in \cite{Salem2024}, AF provides performance comparable to DF at low-to-medium transmitted power levels on the ISLs and outperforms DF at higher transmitted power levels.
Then, We evaluate the ergodic capacity of such a system, considering both THz and FSO ISLs, which introduce ISL noise and pointing error loss to the system.

\subsection{User-Satellite Link Channel Model}
\textcolor{black}{
In this subsection, we describe the flat fading channel model used in our work. For frequencies below $6\,\text{GHz}$, the effects of atmospheric absorption, rain, cloud attenuation, and scintillation on the uplink are assumed to be negligible, as stated in \cite{3GPPrel15}. In a suburban environment, if clutter loss is also negligible, the primary components affecting the signal are free-space path loss (FSPL) and shadow fading\cite{10496842}. The overall path loss can be expressed as:
\begin{equation}
  L= \beta_{\text{up}} X_{sf}  
\end{equation}
where \( X_{sf} \sim \mathcal{N}(0, \sigma_{sf}^2) \) represents the shadow fading effect, modeled as a zero-mean Gaussian random variable with a standard deviation \( \sigma_{sf} \). Values of \( \sigma_{sf} \) for LoS and NLoS conditions in the S-band frequency for various elevation angles are provided in Table 6.6.2-3 of \cite{3GPPrel15}.
The FSPL, denoted as \( \beta_{\text{up}} \), is calculated as\cite{10077720}:
\begin{equation}\label{plup}
    \beta_{\text{up}} = \left( \frac{4 \pi d_m f_{\text{up}}}{c} \right)^2
\end{equation}
where \( d_m \) is the distance between the user and the $m$th satellite, \( f_{\text{up}} \) is the uplink carrier frequency, and \( c = 3 \times 10^8 \, \text{m/s} \) is the speed of light.}

\textcolor{black}{
Due to the dominant LoS component in SatCom systems and the assumption that the user's speed is significantly lower than the satellite's speed, the channel coefficients can be modeled using a Rician distribution, as described in \cite{humadi2024distributed}. The channel coefficient of the $m$th satellite, \( h_m \), is expressed as:
\begin{equation}\label{eq_channel}
h_m = \frac{1}{\sqrt{L_{\text{LoS}}}} \sqrt{\frac{\kappa}{\kappa+1}} h_{\text{LoS}} + \frac{1}{\sqrt{L_{\text{NLoS}}}} \sqrt{\frac{1}{\kappa+1}} h_{\text{NLoS}}.
\end{equation}}In this context, $h_{\text{LoS}}$ represents the LoS channel component, which is stationary, while $h_{\text{NLoS}}\sim\mathcal{CN}(0,1)$ denotes the NLoS multi-path component, following a complex normal distribution. Additionally, since shadow fading differs for LoS and NLoS components, the basic path loss is also distinct for these channels, represented by $L_{\text{LoS}}$ and $L_{\text{NLoS}}$ respectively.
\textcolor{black}{
The parameter \( \kappa \) represents the Rician \( K \)-factor, modeled as a random variable following a complex normal distribution, denoted as \( \mathcal{CN}(\mu_k, \sigma_k) \). This \( K \)-factor is expressed in dB, with its mean \( \mu_k \) and standard deviation \( \sigma_k \) derived from Table 6.7.2-5a in \cite{3GPPrel15}, corresponding to the S-band frequency in a suburban environment.}

\section{Ergodic Capacity analysis with THz ISL}\label{THz ISL}
In this section, we analyze the ergodic capacity for uplink space MIMO communications, considering THz ISLs. \textcolor{black}{During the second hop, all $M$ selected satellites transmit their signals to the MN satellite, through THz ISLs, which are immune to pointing error loss and experience less severe FSPL than FSO ISLs.} The MN satellite receives these signals and performs a weighted summation based on the reliability of each signal. Specifically, signals from weaker channels are assigned smaller weights, while those from stronger channels are given higher weights. This weight vector, denoted as $\mb{v}$, will be further discussed in Subsection \ref{Joint Processing}.
This section begins by describing the ISL channel model. Next, we identify the optimal satellite to serve as the MN satellite. We then outline two specific scenarios for further analysis. Finally, we present the mathematical derivation of the ergodic capacity for each scenario.

\subsection{Inter-Satellite Link Channel Model}
\textcolor{black}{
Turbulence in THz/FSO links occurs primarily in the initial section of the path, limited to the upper layers of the atmosphere. Beyond this region, turbulence does not interfere with signal propagation. Since scintillation effects are caused by turbulent eddies near the Earth's surface, beam scintillation does not affect ISLs. Therefore, unlike uplink and downlink links, ISLs are immune to scintillation effects.
Additionally, atmospheric absorbers and scatterers, such as water vapor, carbon dioxide, and ozone, are mostly concentrated within one to two kilometers of the Earth's surface. As a result, absorption and scattering losses are negligible for ISLs \cite{nie2021channel, maharjan2022atmospheric, 10515772}.
}
In the ISL channel, the attenuation factor decreases consistently with increasing signal frequency. Consequently, the THz and FSO bands experience lower attenuation compared to the RF band \cite{nie2021channel}. Therefore, the main challenges in THz ISL communication are the FSPL and ISL noise, both of which must be carefully considered and modeled.
\textcolor{black}{For FSO ISLs, pointing error loss must also be taken into account, as discussed in Section \ref{FSO pointing error}.}

\textcolor{black}{The received signal at the $n$th satellite from the $m$th satellite through the ISL is expressed as:
\begin{equation}\label{yisl}
    y_{\text{isl}_{mn}} = \frac{\sqrt{p_{\text{isl}}}}{\sqrt{\beta_{\text{isl}_{mn}}}} x_{\text{isl}_m} + n_{\text{isl}_n},
\end{equation}
where:}
\textcolor{black}{\begin{itemize}
    \item $p_{\text{isl}}=P_{T_{\text{isl}}}G_{T_{\text{isl}}}G_{R_{\text{isl}}}$.
    \item $P_{T_{\text{isl}}}$ is the transmit power in the ISL.
    \item $G_{T_{\text{isl}}}$ is the THz/FSO transmit antenna gain.
    \item $G_{R_{\text{isl}}}$ is the THz/FSO receive antenna gain.
    \item $\beta_{\text{isl}_{mn}}$ is the FSPL between satellite $m$ and $n$, calculated as:
    \begin{equation}\label{plisl}
            \beta_{\text{isl}} = \left( \frac{4 \pi d_{\text{isl}_{mn}} f_{\text{isl}}}{c} \right)^2,
    \end{equation}
    where $d_{\text{isl}_{mn}}$ is the distance between satellites $m$ and $n$ in meters, and $f_{\text{isl}}$ is the ISL frequency in Hz.
    \item $x_{\text{isl}_m}$ is the transmitted signal from the $m$th satellite.
    \item $n_{\text{isl}_n}$ is the noise signal at the $n$th satellite, modeled as $n_{\text{isl}_n} \sim \mathcal{CN}(0, \sigma_{n_{\text{isl}}}^2)$. The noise power $\sigma_{n_{\text{isl}}}^2$ is given by:
    \begin{equation}\label{nisl}
          \sigma_{n_{\text{isl}}}^2 = k_B T_{\text{isl}} B_{\text{isl}},
    \end{equation}
    where:
    \begin{itemize}
        \item $k_B = 1.380649 \times 10^{-23}$ is the Boltzmann constant,
        \item $T_{\text{isl}}$ is the noise temperature of the ISL in Kelvin,
        \item $B_{\text{isl}}$ is the ISL bandwidth in Hz.
    \end{itemize}
\end{itemize}
} 

\subsection{MN Satellite Selection}\label{MN Satellite Selection}

As previously noted, in the second hop of transmission, two scenarios are considered. In the first scenario, all satellites share their instantaneous CSI with the MN satellite. In the second scenario, they share statistical CSI. More details are provided in the next subsection. Notably, regardless of whether it is Case 1 or Case 2, the MN satellite always has at least one full instantaneous CSI (its own) and is unaffected by ISL noise. \textcolor{black}{It is preferable to select the satellite with the most reliable channel (highest channel gain) as the MN satellite. According to the channel definition in (\ref{eq_channel}), the channel gain depends on the \( K \)-factor, shadow fading, and FSPL. In a suburban environment, the \( K \)-factor is high enough to approximate $\sqrt{\frac{\kappa}{\kappa+1}} \approx 1$ and $\sqrt{\frac{1}{\kappa+1}} \approx 0$. Additionally, according to 3GPP, the maximum shadow fading difference in the S-band suburban environment is $3.58\,\text{dB}$. However, the minimum difference in user-satellite distances has a significantly larger impact, approximately $39.2\,\text{dB}$. Therefore, the user-satellite distance plays the most critical role in determining channel gain.
However, we aim to select the satellite with the maximum channel gain and the shortest average distance to other satellites. Let the distance from satellite $m$ to satellite $n$ be $d_{\text{isl}_{m,n}}$. The average distance from satellite $m$ to all other satellites is defined as:
\begin{equation}
    \bar{d}_{\text{isl}_m} = \frac{1}{M-1} \sum_{n \neq m} d_{\text{isl}_{m,n}}.
\end{equation}
Our objective is to maximize the channel gain $h_m$ in (\ref{eq_channel}) while minimizing $\bar{d}_{\text{isl}_m}$.
Emphasizing equal weight for channel gain and distance minimization, we have:
\begin{equation}
    \text{index}_{\text{MN}} = \arg \max_{m \in \{1, \ldots, M\}} |h_m-\bar{d}_{\text{isl}_m}|.
\end{equation}
This approach provides a clear mathematical way to identify the satellite with the best balance between maximum channel gain and minimum distance to other satellites.}
Based on the simulation results in Fig. \ref{fig:MN selection}, we select the nearest satellite to the user as the MN satellite for the remainder of the paper. For more details regarding this selection, please refer to the explanation of Fig. \ref{fig:MN selection}.

\subsection{Joint Processing} \label{Joint Processing}
In this section, we introduce a joint processing technique among satellites aimed at minimizing the MSE. Each satellite estimates its own channel, and both the CSI and received signal are shared with the MN satellite. We explore two scenarios: instantaneous CSI sharing and statistical CSI sharing between satellites and the MN satellite. The following subsections provide detailed explanations for each case.

\subsubsection{Case 1: instantaneous  CSI sharing}

In this scenario, all satellites share their data and instantaneous CSI with the MN satellite. The overhead for data sharing is \(\mathcal{O}(M)\) per symbol, while the overhead for CSI sharing is \(\mathcal{O}(M)\) per coherence interval. Before transmitting to the MN satellite, each satellite compensates for the effect of uplink FSPL on the signal. 
\textcolor{black}{Since the locations of the satellites and the center of the user cell are known, this FSPL compensation introduces no additional system load but significantly enhances system performance.}
Subsequently, each satellite \(m\), \(m \in \{1, \dots, M\}\), shares its estimated uplink channel \(\hat{h}_m\) and the modified received signal (i.e., $\sqrt{\beta_{\text{up}_m}}y_{\text{up}_m}$) with the MN satellite.
We assume that channel estimation at each satellite is imperfect, modeled as:
\begin{equation} \label{eq2}
\hat{h}_m = h_m + \tilde{h}_m,
\end{equation}
where \(h_m\) is the \(m\)th uplink channel obtained from (\ref{eq_channel}), and \(\tilde{h}_m \sim \mathcal{CN}(0, \sigma_{\tilde{h}_m}^2), \ m \in \{1, \dots, M\}\) represents the channel estimation error at each satellite. Additionally, the data shared by the \(m\)-th satellite with the MN satellite is affected by additive white Gaussian noise, denoted as \(n_{\text{isl}_m}\).

\textcolor{black}{Using the formula for the received signal in the ISL provided in (\ref{yisl}), the received signal vector at the MN satellite is expressed as:
\begin{equation}\label{ymn}
\mathbf{y}_{\text{MN}} = \sqrt{p_{\text{isl}}} (\bm{\beta}_{\text{up}})^{\frac{1}{2}} \circ (\bm{\beta}_{\text{isl}})^{\frac{-1}{2}} \circ \mathbf{y}_{\text{up}} + \mathbf{n}_{\text{isl}},
\end{equation}
where the symbol \(\circ\) denotes element-wise multiplication. 
\(\bm{\beta}_{\text{up}}\) is an \(M \times 1\) vector, with each element representing the uplink FSPL of a satellite, as defined in (\ref{plup}). Similarly, \(\bm{\beta}_{\text{isl}}\) is an \(M \times 1\) vector, with its elements corresponding to the ISL FSPL of each satellite, as specified in (\ref{plisl}). \(\mathbf{n}_{\text{isl}}\) is an \(M \times 1\) vector whose elements represent ISL noise, modeled as Gaussian zero-mean random variables with variance determined by (\ref{nisl}).
Note that the first elements of \(\bm{\beta}_{\text{isl}}\) and \(\mathbf{n}_{\text{isl}}\) are set to \(1\) and \(0\), respectively, because the first satellite is designated as the MN satellite, as explained in Section \ref{MN Satellite Selection}.}

\textcolor{black}{To enhance readability, we define an \(M \times 1\) vector as follows:
\begin{equation}\label{c}
    \mathbf{c} = \sqrt{p_{\text{isl}}}(\bm{\beta}_{\text{up}})^{\frac{1}{2}} \circ (\bm{\beta}_{\text{isl}})^{\frac{-1}{2}},
\end{equation}
and rewrite (\ref{ymn}) as:
\begin{equation}\label{ymn2}
    \mathbf{y}_{\text{MN}} = \text{diag}(\mathbf{c}) \mathbf{y}_{\text{up}} + \mathbf{n}_{\text{isl}}.
\end{equation}
In (\ref{ymn2}), \(\mathbf{y}_{\text{up}}\) represents the received uplink signal vector, expressed as:
\begin{equation}
    \mathbf{y}_{\text{up}} = \sqrt{p}s\mathbf{h} + \mathbf{n}_{\text{up}},
\end{equation}
where \(p = P_T G_T G_R\), with \(P_T\) denoting the transmit power, \(G_T\) the gain of the transmit antenna, and \(G_R\) the gain of the satellite's receive antenna. \(\mathbf{n}_{\text{up}}\) is an \(M \times 1\) vector, where each element \(n_{\text{up}_m} \sim \mathcal{CN}(0, \sigma_{n_{\text{up}}}^2)\) represents Gaussian noise. The noise power is calculated as, $\sigma_{n_{\text{up}}}^2 = k_B \times T_{\text{up}} \times B_{\text{up}}$,
where \(T_{\text{up}} = 290 \, \text{K}\) is the uplink noise temperature, and \(B_{\text{up}}\) is the uplink bandwidth in Hz.
Furthermore, \(\mathbf{h} = \mathbf{\hat{h}} - \mathbf{\tilde{h}}\), where:
\begin{itemize}
    \item \(\mathbf{h} = [h_1, \dots, h_M]^T\): the uplink channel vector,
    \item \(\mathbf{\hat{h}} = [\hat{h}_1, \dots, \hat{h}_M]^T\): the estimated channel vector,
    \item \(\mathbf{\tilde{h}} = [\tilde{h}_1, \dots, \tilde{h}_M]^T\): the channel estimation error vector.
\end{itemize}}

Now, assuming the joint detection vector is represented by \( \mathbf{v} \in \mathbb{C}^{1 \times M} \), the detected signal can be expressed as:
\begin{align}
\hat{s} &= \mathbf{v} \mathbf{y}_{\text{MN}} = \sqrt{p} s \mathbf{v} \text{diag}(\mathbf{c}) \mathbf{h} + \mathbf{v} \text{diag}(\mathbf{c}) \mathbf{n}_{\text{up}} + \mathbf{v} \mathbf{n}_{\text{isl}} \nonumber \\
&= \sqrt{p} s \mathbf{v} \text{diag}(\mathbf{c}) (\mathbf{\hat{h}} - \mathbf{\tilde{h}}) + \mathbf{v} \text{diag}(\mathbf{c}) \mathbf{n}_{\text{up}} + \mathbf{v} \mathbf{n}_{\text{isl}}. \label{EQ5}
\end{align}
The objective is to determine the detection vector that minimizes the mean squared error (MSE). This can be formulated as the following optimization problem:
\begin{equation}
\begin{array}{cl}
\min\limits_{\mathbf{v}} & \mathbb{E}\left\{\left|\hat{s} - s\right|^2\right\},
\end{array}
\end{equation}
where the expectation is taken over the additive noise at each satellite, the noise at the ISL, the channel estimation error, and the transmitted symbol.
To find the optimal solution, the first-order optimality condition is applied. Specifically, the gradient of the objective function with respect to the detection vector must be zero at the optimal point. By setting the gradient of the objective function to zero and solving for the detection vector, the optimal detection vector is obtained as:

\begin{align}
    \mathbf{v} = & \ \sqrt{p}\hat{\mathbf{h}}^H\text{diag}(\mathbf{c}) \left( p(\text{diag}(\mathbf{c}))^2\hat{\mathbf{h}}\hat{\mathbf{h}}^H \right.\\ \nonumber 
    & \left. + \ p(\text{diag}(\mathbf{c}))^2\mathbf{\Sigma}_{\tilde{h}} + \mathbf{\Sigma}_{n_{\text{isl}}} + \sigma_{n_{\text{up}}}^2(\text{diag}(\mathbf{c}))^2 \right)^{-1},
\end{align}
where 
\begin{equation}
    \mathbf{\Sigma}_{\tilde{h}} = \mathbb{E}\left\{\tilde{\mathbf{h}}\tilde{\mathbf{h}}^H\right\} = \text{diag}\left[\sigma_{\tilde{h}_1}^2,\ldots,\sigma_{\tilde{h}_L}^2\right],
\end{equation}
and
\begin{align}
    \mathbf{\Sigma}_{n_{\text{isl}}} &= \mathbb{E}\left\{\mathbf{n}_{\text{isl}}\mathbf{n}_{\text{isl}}^H\right\} = \text{diag}\left[0,\sigma_{n_{\text{isl}_2}}^2,\ldots,\sigma_{n_{\text{isl}_M}}^2\right].
\end{align}
The capacity in this case is given by the mutual information between \( s \) and \( \hat{s} \), expressed as:
\begin{equation}\label{mutual information}
    I(s;\hat{s}) = H(\hat{s}) - H(\hat{s}|s),
\end{equation}
where $H(\hat{s})\leq \log(\pi e (p\mb{v}(\text{diag}(\mb{c}))^2\mb{h}\mb{h}^H\mb{v}^H+\sigma_{n_{\text{up}}}^2\mb{v}(\text{diag}(\mb{c}))^2\mb{v}^H+\mb{v}\mb{\Sigma}_{n_{\text{isl}}}\mb{v}^H))$ \cite{duman2008coding},   and $H(\hat{s}|s)=\log(\pi e (\sigma_{n_{\text{up}}}^2\mb{v}(\text{diag}(\mb{c}))^2\mb{v}^H+\mb{v}\mb{\Sigma}_{n_{\text{isl}}}\mb{v}^H))$.
To obtain the capacity, we assume \( s \) follows a Gaussian distribution. Then \( \hat{s} \), as in (\ref{mutual information}), would also follow a Gaussian distribution. The capacity for the full CSI scenario is given by: 
\begin{equation}\label{rate FC}
    R_{\text{FC}} = \log \left( 1+\frac{p\mathbf{v}(\text{diag}(\mathbf{c}))^2\mathbf{h}\mathbf{h}^H\mathbf{v}^H}{\sigma_{n_{\text{up}}}^2\mathbf{v}(\text{diag}(\mathbf{c}))^2\mathbf{v}^H+\mathbf{v}\mathbf{\Sigma}_{n_{\text{isl}}}\mathbf{v}^H} \right).
\end{equation}
This capacity is maximized when there is perfect CSI, i.e., \( \hat{\mathbf{h}} = \mathbf{h} \). However, with uncertain estimated CSI, the achievable rate decreases. This is demonstrated in Section \ref{Numerical Results} with Fig. \ref{fig:1} and Fig. \ref{fig:2}.

\subsubsection{Case 2: statistical CSI sharing}
In this scenario, unlike Case 1, the instantaneous estimated CSI of each satellite is not shared with the MN satellite. Therefore, each satellite $m$, $m \in \{1, \ldots, M\}$, estimates its own channel imperfectly and can perform local processing to mitigate the effects of channel attenuation in the uplink ($\hat{y}_{\text{up}_m}=\frac{y_{\text{up}_m}}{\hat{h}_m}$).
Taking into account the estimated channel in (\ref{eq2}), the local processed signal in the uplink would be,
\begin{equation}
    \hat{y}_{\text{up}_m}=\frac{\sqrt{p}h_ms+n_{\text{up}_m}}{\hat{h}_m}=\sqrt{p}s-\sqrt{p}s\frac{\tilde{h}_m}{\hat{h}_m}+\frac{n_{\text{up}_m}}{\hat{h}_m}.
\end{equation}
Subsequently, each satellite $m \in \{2, \ldots, M\}$ send its processed signal $\hat{y}_{\text{up}_m}$ to the MN satellite along with the following long-term information: (i) the average power of the estimated channel $\mathbb{E}\{|\hat{h}_m|^2\}$, (ii) the average power of the estimation error $\mathbb{E}\{|\tilde{h}_m|^2\}$, and (iii) the average power of the inverse estimated channel $\mathbb{E}\{|\frac{1}{\hat{h}_m}|^2\}$. Since this information is long-term, sharing it with the MN satellite does not increase the system overhead. Consequently, in this case, the system overhead is of the order $\mathcal{O}(M)$. After each satellite completes local processing, the processed data \textcolor{black}{is affected by the ISL FSPL} and combined with ISL noise and sent to the MN satellite for central processing.

The central processing vector is denoted by $\mb{v} \in \mathbb{C}^{1 \times M}$, and the final detected signal at the MN satellite is given by $\hat{s} = \mb{v} \mb{y}_{\text{MN}}$. Similar to Case 1, the following optimization problem is solved,
\begin{equation}\label{optimization 2}
    \begin{array}{cl}
    \min\limits_{\mb{v}}&  \mathbb{E}\{|\hat{s}-s|^2\},
    \end{array}
\end{equation}
where the expectation is taken over the additive noise, channel estimation error, the transmitted symbol, and the estimated uplink channel at each satellite. 
Given the distribution of the channel estimation error, the objective function in (\ref{optimization 2}) can be rewritten as:
\begin{align}
    \mathbb{E}\{|\hat{s}-s|^2\}=&p\mb{v}(\text{diag}(\mb{c}_2))^2\mb{u}\mb{u}^T\mb{v}^H-\sqrt{p}\mb{v}\text{diag}(\mb{c}_2)\mb{u}\nonumber\\&-\sqrt{p}\mb{u}^T\text{diag}(\mb{c}_2)\mb{v}^H+\mb{v}(\text{diag}(\mb{c}_2))^2\mb{B}\mb{v}^H\nonumber\\&+p\mb{v}(\text{diag}(\mb{c}_2))^2\mb{S}\mb{v}^H+\mb{v}\mb{\Sigma}_{n_{\text{isl}}}\mb{v}^H+1,
\end{align}
where $\mb{u}$ represents an all-one $M\times 1$ vector. \textcolor{black}{$\mb{c}_2=\text{diag}(\sqrt{p_{isl}}(\bm{\beta}_{isl})^{\frac{-1}{2}})$}. 
The matrix $\mb{B}$ is a diagonal $M\times M$ matrix with its $m$th diagonal entry ($m\neq 1$), equal to $\sigma_{n_{\text{up}}}^2\mathbb{E}\{|\frac{1}{\hat{h}_{m}}|^2\}$, while its first diagonal entry is $\frac{\sigma_{n_{\text{up}}}^2}{|\hat{h}_1|^2}$.v. 
The matrix $\mb{S}$ is also an $M\times M$ diagonal matrix with its  $m$th diagonal entry ($m\neq 1$), equal to $\mathbb{E}\{|\tilde{h}_{m}|^2\}\mathbb{E}\{|\frac{1}{\hat{h}_{m}}|^2\}$, while its first diagonal entry is $\mathbb{E}\{|\Tilde{h}_1|^2\}|\frac{1}{\hat{h}_1}|^2$. In other words, $\mb{B}$ and $\mb{S}$ can be respectively represented by:
\begin{align}
    \mb{B}=\diag\Bigg[&\frac{\sigma_{n_{\text{up}}}^2}{|\hat{h}_1|^2}, \sigma_{n_{\text{up}}}^2\mathbb{E}\{\bigg|\frac{1}{\hat{h}_{2}}\bigg|^2\},...,\sigma_{n_{\text{up}}}^2\mathbb{E}\{\bigg|\frac{1}{\hat{h}_{M}}\bigg|^2\}
    \Bigg],
\end{align}
and 
\begin{align}
    \mb{S}=\diag\Bigg[&\mathbb{E}\{|\Tilde{h}_1|^2\}\bigg|\frac{1}{\hat{h}_1}\bigg|^2, \mathbb{E}\{|\tilde{h}_{2}|^2\}\mathbb{E}\{\bigg|\frac{1}{\hat{h}_{2}}\bigg|^2\},...,\\ \nonumber
    &\mathbb{E}\{|\tilde{h}_{M}|^2\}\mathbb{E}\{\bigg|\frac{1}{\hat{h}_{M}}\bigg|^2\}\Bigg].
\end{align}

Now, by using the first-order optimality condition, the optimal solution for the optimization problem in (\ref{optimization 2}) is given by: 
\begin{align}\label{detection, case2}    
    \mb{v}&=\sqrt{p}\,\text{diag}(\mb{c}_2)\mb{u}^T\bigg(p(\text{diag}(\mb{c}_2))^2\mb{u}\mb{u}^T+p(\text{diag}(\mb{c}_2))^2\mb{S}\nonumber\\&+(\text{diag}(\mb{c}_2))^2\mb{B}+\mb{\Sigma}_{n_{\text{isl}}}\bigg)^{-1}.
\end{align}
The capacity in this scenario is given by the mutual information between the transmitted signal $s$ and the final centrally-processed signal $\hat{s}$, as in (\ref{mutual information}), where
$H(\hat{s})\leq \log(\pi e (p\mb{v}(\text{diag}(\mb{c}_2))^2\mb{u}\mb{u}^T\mb{v}^H+\mb{v}(\text{diag}(\mb{c}_2))^2\mb{B}\mb{v}^H+\mb{v}\mb{\Sigma}_{n_{\text{isl}}}\mb{v}^H))$ and $H(\hat{s}|s)=\log(\pi e( \mb{v}(\text{diag}(\mb{c}_2))^2\mb{B}\mb{v}^H+\mb{v}\mb{\Sigma}_{n_{\text{isl}}}\mb{v}^H))$. Thus, the achievable rate for the partial CSI case can be written as:
\begin{equation}\label{rate PC}
R_{\text{PC}}=\log \left( 1+\frac{p\mb{v}(\text{diag}(\mb{c}_2))^2\mb{u}\mb{u}^T\mb{v}^H}{\mb{v}(\text{diag}(\mb{c}_2))^2\mb{B}\mb{v}^H+\mb{v}\mb{\Sigma}_{n_{\text{isl}}}\mb{v}^H} \right).
\end{equation}
This capacity is maximized when perfect CSI is available, i.e., $\mb{\hat{h}}=\mb{h}$. For estimated values of CSI, the given rate is reduced. 
\vspace{-10.5mm}
\subsection{Ergodic Capacity Analysis}
In this section, we assume that the CSI distribution is known to each satellite. As both the number and positions of the satellites change over time, our focus is on determining the ergodic capacity of the distributed space MIMO system. In the second hop, the instantaneous CSI is shared with the MN satellite. The ergodic capacity for uplink transmission from a single user to $M$ cooperating satellites is defined by:
\begin{equation}
    C_e = \mathbb{E}\left\{\log_2(1 + \rho \mb{h}^H \mb{h})\right\},
\end{equation}
where $\rho = \frac{p\, \bigg(p_{isl}\, \frac{\beta_{\text{up}}}{\beta_{\text{isl}}}\bigg)}{\bigg(p_{isl}\, \frac{\beta_{\text{up}}}{\beta_{\text{isl}}}\bigg)\, \sigma_{n_{\text{up}}}^2 + \sigma_{n_{\text{isl}}}^2}$ represents the channel SNR and the expectation is taken over the random vector $\mb{h}$. In this context, we assume that the channel follows a complex Rician distribution, denoted as $\mb{h} \sim \mathcal{CN}(\mb{m}_h, \mb{\Sigma}_h)$, where $\mb{m}_h \in \mathbb{C}^{M \times 1}$ is the mean vector, and $\mb{\Sigma}_h$ is the covariance matrix.
Given the distances between satellites, it is reasonable to infer that the channels are independent, allowing the covariance matrix to be represented as a diagonal matrix. Additionally, in a dense satellite constellation, it is assumed that the path loss for all channel coefficients is nearly identical, resulting in approximately the same channel variances between the user and each satellite. Therefore, we have $\mb{\Sigma}_h = \sigma_h^2 \mb{I}$. Note that the channels are independently and identically distributed between each user and the satellites. Also, we define $w = \mb{h}^H \mb{h}$, which follows a non-central Wishart distribution with the following PDF \cite{ratnarajah2003topics},
\begin{equation}
    f(w)=\frac{e^{-\Omega}}{\Gamma(M)(\sigma_h^2)^{M}}e^{\frac{-w}{\sigma_h^{2}}}w^{M-1}\ _0F_1(M,\frac{\Omega w}{\sigma_h^2}),
\end{equation}
where $\Omega$ is represented by $\Omega=\frac{\mb{m}^H\mb{m}}{\sigma_h^2}$, $\Gamma(M)=(M-1)!$ denotes the gamma function and  $_0F_1(b,x)$ is the generalized hypergeometric function \cite{james1964distributions} defined by: 
\begin{equation} \label{hypergeometric}
    _0F_1(b,x) = \sum_{k=0}^{\infty}\frac{x^k}{(b)_kk!}.
\end{equation}
The hypergeometric coefficient in (\ref{hypergeometric}) is defined as $(b)_k = b(b+1)\cdots(b+k-1)$.
The ergodic capacity can be reformulated by the eigenvalue of $\mb{h}\mb{h}^H$ as (see Theorem 7.10 of \cite{ratnarajah2003topics}):
\begin{align}\label{Ergodic Capacity}
    C_e &= \int_0^{\infty}\log_2(\rho w + 1)f(w) \, \text{d}w = \frac{\Psi e^{-\Omega}}{(M-1)!(\sigma_h^2)^{M}},
\end{align}
where $\Psi = \int_0^{\infty}\log_2(\rho w + 1)e^{\frac{-w}{\sigma_h^{2}}}w^{M-1}\ _0F_1(M, \frac{\Omega w}{\sigma_h^2}) \, \text{d}w$.
We use $\mb{Lemma \ 1}$ in Appendix \ref{FirstAppendix}, to introduce a closed-form approximation for the intergral in (\ref{Ergodic Capacity}) as:

\begin{align}\label{Capacity approximation}
    C_e \approx \frac{\rho \sigma_h^2 e^{-\Omega} M^{M+2}}{(M - \Omega)^{M+1}}.
\end{align}
It is important to note that the equation in (\ref{Capacity approximation}) holds only if $M > \Omega$.

\section{Ergodic Capacity analysis with FSO ISL}\label{FSO ISL}

In this section, we explore a scenario of space MIMO communications where all satellites simultaneously transmit their received signals to the MN satellite during the second hop.
\textcolor{black}{
As explained in section \ref{THz ISL}, unlike uplink and downlink, in ISLs the scintillation effects, absorption, scattering, and attenuation can be neglected.}
Consequently, our analysis focuses on FSO ISLs, incorporating ISL noise, ISL FSPL, and ISL pointing error loss into our mathematical model to accurately capture the complexities of FSO communications in the space environment.
The received signal vector at the MN satellite, considering compensation for uplink FSPL by each satellite, can be represented by:
\textcolor{black}{
\begin{align}\label{eq31}
y_{\text{MN}}&=\sqrt{p}\sqrt{p_{\text{isl}}} \, \bm{\alpha}_p \bigg((\bm{\beta}_{\text{up}})^{\frac{1}{2}}\circ (\bm{\beta}_{\text{isl}})^{\frac{-1}{2}}\circ \mb{h}\bigg)s+ \nonumber \\ &\sqrt{p_{\text{isl}}} \, \bm{\alpha}_p \bigg((\bm{\beta}_{\text{up}})^{\frac{1}{2}}\circ (\bm{\beta}_{\text{isl}})^{\frac{-1}{2}}\circ \mb{n}_{\text{up}}\bigg) + \mb{u}_2\mb{n}_{\text{isl}},
\end{align}
where, $\bm{\alpha}_p \in \mathcal{C}^{1 \times M}$ denotes the ISL pointing error loss vector, influencing the communication between the satellites and the MN satellite. Additionally, $\mathbf{u}_2$ represents a $1 \times M$ all-ones vector. For simplicity, we substitute $\mathbf{c}$ from~\eqref{c} into~\eqref{eq31} and rewrite it as:
\begin{equation}
    y_{\text{MN}} = \sqrt{p} \, \bm{\alpha}_p \, \text{diag}(\mathbf{c}) \mathbf{h} \, s + \bm{\alpha}_p \, \text{diag}(\mathbf{c}) \mathbf{n}_{\text{up}} + \mathbf{u}_2 \mathbf{n}_{\text{isl}}.
\end{equation}} \vspace{-10mm}

\subsection{FSO Pointing Error Loss Model}\label{FSO pointing error}
\textcolor{black}{
The pointing error loss caused by misalignment is a result of the displacement of the laser beam in both the elevation and azimuth directions. These displacements are modeled as zero-mean, independent Gaussian random variables. Consequently, the radial displacement at the receiver, $\theta$, is determined as the root sum of squares of the elevation and azimuth components, leading to a Rayleigh distribution,} $f_{\theta}(\theta)=\frac{\theta}{\sigma_p^2}e^{\frac{-\theta^2}{2\sigma_p^2}}$ \cite{9108615, 7553489}. Where $\sigma_p$ is the standard deviation of the radial displacement at the receiver.
In FSO communications, the normalized collected power at the receiver can be estimated as\cite{9822386}:
\begin{equation}\label{eq32}
   {\alpha}_{p}(\theta)\approx \text{exp}\left(\frac{-2\theta^2}{\omega_{z}^2}\right),
\end{equation}
Due to diffraction, the optical beam undergoes angular spreading and reaches the receiver with a beam width denoted by:
\begin{equation}
    \omega_d = \omega_0 \sqrt{1 + \left(\frac{d_{\text{isl}}}{d_R}\right)^2},
\end{equation}
where $d_{\text{isl}}$ represents the ISL distance, and $d_R = \frac{\pi \omega_0^2}{\lambda_{\text{isl}}}$ is defined as the Rayleigh range. The initial beam width at $d_{\text{isl}}=0$ is given by $\omega_0$ and $\lambda_{\text{isl}}$ is the wavelength of the optical signal. The comprehensive beam width accounting for both diffraction and turbulence effects is expressed as \cite{aladeloba2013optically}:
\begin{equation}\label{eq33}
    \omega_z = \omega_d \sqrt{1 + 1.33 \, \sigma_R^2 \left(\frac{2\,d_{\text{isl}}}{k\, \omega_d^2}\right)^{\frac{5}{6}}},
\end{equation}
where $k = \frac{2\pi}{\lambda_{\text{isl}}}$ denotes the optical wave number, and $\sigma_R^2 = 1.23 , C_n^2 , k^{\frac{7}{6}} , d_{\text{isl}}^{\frac{11}{6}}$ represents the Rytov variance. In conditions of low turbulence, which are common in ISLs~\cite{nie2021channel}, $C_n^2 \ll 1$, resulting in $\sigma_R^2 \ll 1$. As a result, $\omega_z \approx \omega_d$.
Given a transformation $y = G(x)$ where $G$ is a monotonic function, then
$f_y(y) = f_x(G^{-1}(y)) \left| \frac{d(G^{-1}(y))}{dy} \right|$.
Using this relationship and equation (\ref{eq32}), we can determine the PDF of the normalized irradiance affected by pointing errors, \( \alpha_{p} \), as \cite{9535285}:
\begin{equation}\label{eq34}
  f_{\alpha_{p}}(\alpha_{p})= \gamma^2 \alpha_{p}^{\gamma^2-1}. \quad \quad \quad 0\leq \alpha_{p} \leq 1
\end{equation}
The ratio $\gamma=\frac{\omega_z}{2 \sigma_p}$ represents the relationship between the equivalent beamwidth and the standard deviation of the Gaussian beam's displacement due to pointing errors. This parameter is crucial as it quantifies the severity of the pointing error's impact on the system \cite{9108615}.

\subsection{Approximated Upper Bound of the Ergodic Capacity}
We assume that neither the user nor the satellites possess CSI, while the MN satellite has complete knowledge of both $\bm{\alpha}_{p}$ and $\bm{\alpha}_{p}\mb{h}$. According to the given model, the ergodic capacity (measured in bits/s/Hz) of the system is described as follows \cite{5452205}:
\begin{equation}\label{eq35}
C_e=E\{\text{log}_2(1+R_sR_n^{-1})\},
\end{equation}
where $R_s$ and $R_n$ are given by: 
\begin{equation}\label{eq36}
R_s=p\,\bm{\alpha}_{p}\bigg(\text{diag}(\mb{c})\bigg)^2\mb{h}\mb{h}^H\bm{\alpha}_{p}^H,
\end{equation}
and
\begin{equation}\label{eq37}
R_n=\sigma_{n_{\text{up}}}^2\bm{\alpha}_{p}\bigg(\text{diag}(\mb{c})\bigg)^2\bm{\alpha}_{p}^H+M\sigma_{n_{\text{isl}}}^2.
\end{equation}
According to Jensen's inequality, for a random variable $X$ and a concave function $F$, it holds that $E\{F(X)\} \leq F(E\{X\})$ \cite{mcshane1937jensen}. Applying this principle and incorporating (\ref{eq36}) and (\ref{eq37}) into (\ref{eq35}), we can express the upper bound of (\ref{eq35}) as follows:
\begin{align}\label{eq38}
& E\left\{\text{log}_2\left(1+\frac{p\Phi\Phi^H}{\sigma_{n_{\text{up}}}^2\bm{\psi}\bm{\psi}^H+M\sigma_{n_{\text{isl}}}^2}\right)\right\}\leq\\
& \nonumber \text{log}_2\left(1+E\left\{\frac{p\Phi\Phi^H}{\sigma_{n_{\text{up}}}^2\bm{\psi}\bm{\psi}^H+M\sigma_{n_{\text{isl}}}^2} \right\} \right),
\end{align}
where $\Phi=\bm{\alpha}_{p}\,\text{diag}(\mb{c})\,\mb{h}$ and $\bm{\psi}=\bm{\alpha}_{p}\,\text{diag}(\mb{c})$. Then we obtain,
\begin{equation}\label{Psi}
    \bm{\psi}\bm{\psi}^H=p_{\text{isl}}\,\displaystyle\sum_{i=1}^{M} \frac{\beta_{\text{up}_i}}{\beta_{\text{isl}_i}}\, \alpha_{p_i}^2,
\end{equation}
and
\begin{align}\label{Phi}
  \Phi\Phi^H &= p_{\text{isl}}\, \bigg( \displaystyle\sum_{i=1}^{M} \frac{\beta_{\text{up}_i}}{\beta_{\text{isl}_i}}\, \alpha_{p_i}^2 |h_i|^2+ \nonumber \\& \displaystyle\sum_{i=1}^{M}\displaystyle\sum_{\substack{j=1 \\ j \neq i}}^{M} \frac{\sqrt{\beta_{\text{up}_i}} \, \sqrt{\beta_{\text{up}_j}} }{\sqrt{\beta_{\text{isl}_i}} \, \sqrt{\beta_{\text{isl}_j}}} \alpha_{p_i} \alpha_{p_j}   h_ih_j^* \bigg).
\end{align}

\textcolor{black}{We then assume a dense satellite constellation where all user-satellite distances are nearly identical, leading to approximately the same uplink FSPL between the user and each satellite. In such systems, the equations in (\ref{Psi}) and (\ref{Phi}) can be further simplified by additionally assuming equal ISL distances for all satellites.} Then, we divide both nominator and denominator of the (\ref{eq38}) by $M$. Utilizing the law of large numbers, and considering that both $\bm{\alpha_p}$ and $\mb{h}$ have independently and identically distributed (i.i.d.) elements and are independent of each other, we can state that for a large number of satellites $M$, (\ref{Phi}) could be rewritten as:\textcolor{black}{
{\small
\begin{equation}
\begin{aligned}
    \Phi \Phi^H = p_{\text{isl}} \left( \frac{{\beta}_{\text{up}}}{{\beta}_{\text{isl}}} \right) & 
    \Bigg[
         \frac{1}{M} \sum_{i=1}^M \alpha_{p_i}^2 |h_i|^2  \\
        & + \frac{1}{M} \sum_{i=1}^M \sum_{\substack{j=1 \\ j \neq i}}^{M} \alpha_{p_i} \alpha_{p_j} h_i h_j^* 
    \Bigg]_{M \to \infty} \\
   \!\! = p_{\text{isl}}\!\left(\!\frac{{\beta}_{\text{up}}}{{\beta}_{\text{isl}}} \right)^2\! &
    \Bigg(\!
        \left( \mathbb{E}\{\alpha_p^2\} \mathbb{E}\{|h|^2\} \right)\! +\! M \left(\mathbb{E}\{\alpha_p\}\right)^2 \left(\mathbb{E}\{h\}\right)^2\!
    \Bigg)\!,
\end{aligned}
\end{equation}
}
and
\begin{align}
    \bm{\psi}\bm{\psi}^H =& p_{\text{isl}} \left( \frac{{\beta}_{\text{up}}}{{\beta}_{\text{isl}}} \right) \frac{1}{M} \sum_{i=1}^M \alpha_{p_i}^2 \bigg|_{M \to \infty} =\\
    & \nonumber p_{\text{isl}} \left( \frac{{\beta}_{\text{up}}}{{\beta}_{\text{isl}}} \right) \mathbb{E}\{\alpha_{p}^2\}.    
\end{align}
} 
The expected values $\mathbb{E}\{\alpha_{p}\}$ and $\mathbb{E}\{\alpha_{p}^2\}$ can be obtained using the PDF of \( \alpha_{p} \), in (\ref{eq34}) as:
\begin{equation}\label{eq40}
\mathbb{E}\{{\alpha}_{p}\} = \frac{\omega_{z^2}}{\omega_{z}^2+4\sigma_p^2}=\frac{\gamma^2}{\gamma^2+1},
\end{equation}
and
\begin{equation}\label{eq41}
\mathbb{E}\{{\alpha}_{p}^2\} = \frac{\omega_{z}^2}{\omega_{z}^2+8\sigma_p^2}=\frac{\gamma^2}{\gamma^2+2}.
\end{equation}
Substituting the derived results into (\ref{eq38}), the approximated upper bound of the Ergodic capacity can be expressed as follows:

\begin{align}\label{eq42}
& C_{e_{max}} \approx\text{log}_2(1+\Psi),
\end{align} 
where,
$\Psi=\frac{p\,p_{\text{isl}}\frac{{\beta}_{\text{up}}}{{\beta}_{\text{isl}}} \bigg(\frac{\gamma^2}{\gamma^2+2} \mathbb{E}\{|h|^2\} + M (\frac{\gamma^2}{\gamma^2+1})^2 \mathbb{E}\{h\}^2\bigg)}{\sigma_{n_{\text{up}}}^2 \, p_{\text{isl}}\frac{{\beta}_{\text{up}}}{{\beta}_{\text{isl}}} \frac{\gamma^2}{\gamma^2+2} + \sigma_{n_{\text{isl}}}^2}$.

\section{Numerical Results}\label{Numerical Results}

\begin{figure}
    \centering
    \includegraphics[scale=0.5]{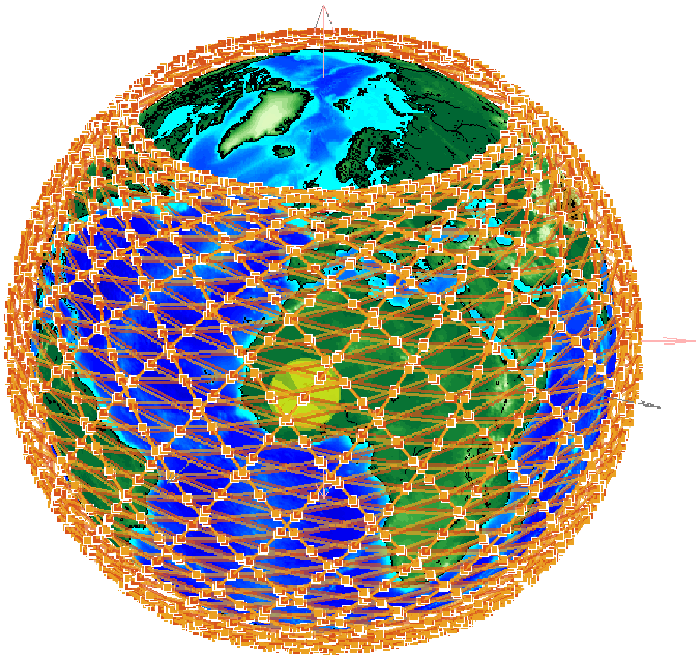}
    \caption{Our employed constellation with Starlink specifications.}
    \label{fig:Constellation}
\end{figure}

In this section, we present the results for both THz and FSO ISL cases. The simulations are based on the Starlink constellation, utilizing two distinct layers corresponding to two groups from the Starlink launch plans \cite{liang2021phasing}.
The first group consists of 1,584 satellites at an altitude of \(540\,\text{km}\) with a \(53.2^\circ\) inclination, while the second group comprises 1,584 satellites at an altitude of \(550\,\text{km}\) with a \(53^\circ\) inclination. Additional specifications of the constellation are provided in Table~\ref{tab1}. A total of 3,168 satellites are included in the constellation, as illustrated in Fig.~\ref{fig:Constellation}.

\begin{table}[h!]
\centering
\caption{Parameters and Specifications}
\label{tab1}
\begin{tabular}{@{}ll@{}}
\toprule
\textbf{Parameter}                          & \textbf{Value}                    \\ \midrule
\multicolumn{2}{c}{\textbf{Satellite Constellation Specification}} \\ \midrule
Number of orbital planes                   & 22 \cite{liang2021phasing}                               \\
Number of satellites per plane             & 72 \cite{liang2021phasing}                               \\
Satellite altitudes                        & 540, 550 km                       \\
RAAN spread                                & $36^\circ$                        \\
Inclination angle                          & $53.2^\circ$, $53^\circ$          \\
True anomaly phasing                       & $1.1364^\circ$                    \\
Minimum elevation angle                    & $30^\circ$                         \\ \midrule
\multicolumn{2}{c}{\textbf{User-Satellite Link}} \\ \midrule
Mobile phone transmit power ($P_T$)        & -6\,\text{dBW} \cite{3GPPrel15}                           \\
Mobile phone transmit antenna gain ($G_T$) & 5\,\text{dBi}      \cite{3GPPrel15}                        \\
Satellite receive antenna gain ($G_R$)     & 35\,\text{dBi}                            \\
Uplink signal bandwidth ($B_{\text{up}}$)         & 20\,\text{MHz}     \cite{3GPPrel15}                        \\
Uplink frequency band ($f_{\text{up}}$)           & 2\,\text{GHz}      \cite{3GPPrel15}                        \\
Uplink ambient noise temperature ($T_{\text{up}}$) & 290\,\text{K}                                               \\
Shadow fading standard deviation ($\sigma_{sf}$) & Table 6.6.2-3 of \cite{3GPPrel15}          \\
Mean and standard deviation of \( K \)-factor ($\mu_k$, $\sigma_k$) & Table 6.7.2-5a of \cite{3GPPrel15}\\
Environment type                                & Suburban                          \\ \midrule
\multicolumn{2}{c}{\textbf{Inter-Satellite Link}} \\ \midrule
Satellite transmit power ($P_{T_{\text{isl}}}$)   & 5\,\text{dBW} \cite{Salem2024}                             \\
Satellite THz/FSO transmit antenna gain ($G_{T_{\text{isl}}}$) & 60\,/\,90\,\text{dBi} \cite{Salem2024}\\
Satellite THz/FSO receive antenna gain ($G_{R_{\text{isl}}}$)  & 60\,/\,90\,\text{dBi} \cite{Salem2024}\\
ISL ambient noise temperature ($T_{\text{isl}}$)  & 7000\,\text{K}     \cite{Salem2024}\\
THz/FSO ISL frequency band ($f_{\text{isl}}$)             & 1\,/\,193\,\text{THz}                             \\
ISL bandwidth ($B_{\text{isl}}$)             & $0.02 f_{\text{isl}}$  \cite{nie2021channel}                           \\ \bottomrule

\end{tabular}
\end{table}

\textcolor{black}{
For uplink communication, the S-band frequency is adopted with a carrier frequency of $f_{\text{up}} = 2\,\text{GHz}$ and a bandwidth of $B_{\text{up}} = 20\,\text{MHz}$, as specified in \cite{3GPPrel15}. The transmitter is assumed to be a mobile phone. The main simulation parameters for the uplink are listed in Table \ref{tab1}.
The users are located in the Lake District National Park, UK, where terrestrial infrastructure is limited, particularly in off-the-grid areas. The simulation considers users within a $40\,\text{km}$ radius of a point at latitude $54.526\degree$ and longitude $-3.3\degree$ \cite{omid2024reinforcement}.
At least $20$ satellites from our constellation are within the user's visibility region at the specified location. The channel estimation error is modeled as $\sigma_{\tilde{h}_m}^2 = \epsilon^2 \text{var}(h_m)$, where $\epsilon = 0.05$. Unless otherwise stated, all simulations for THz ISLs are conducted with a $1\,\text{THz}$ ISL carrier frequency, and the ISL bandwidth is defined as $B_{\text{isl}} = 0.02f_{\text{isl}}$.} 

For the ergodic capacity evaluation of FSO communications with pointing error loss, we consider a transmitter beam radius $\omega_0 = 0.05\, \text{m}$, and $\gamma=1.1$ \cite{9108615}. Additionally, we assume the ISL carrier frequency to be $193\,\text{THz}$, as the $1550\,\text{nm}$ or $193\,\text{THz}$ band is especially preferred for free space transmission due to its lower attenuation levels and the availability of high-quality transmitter and detector components that are suitable for Wavelength Division Multiplexing (WDM) operations \cite{alkholidi2014fso}.

\begin{figure}
   \centering
   \includegraphics[scale=0.34]{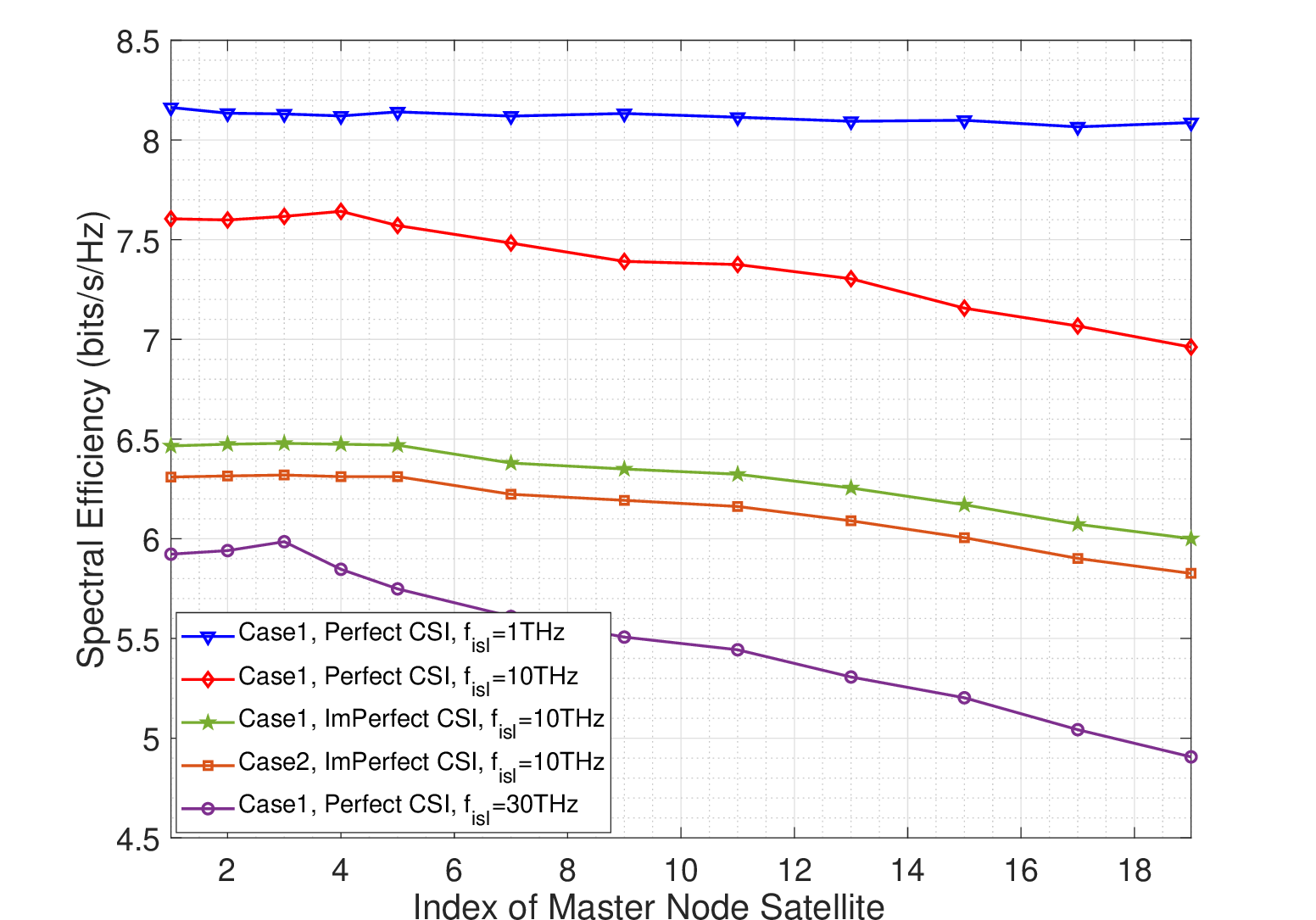}
   \caption{The effect of MN satellite selection on the spectral efficiency.}
    \label{fig:MN selection}
\end{figure}

\textcolor{black}{In Fig. \ref{fig:MN selection}, we illustrate which satellite is suitable to serve as the MN satellite, responsible for processing the received signals from all satellites while considering ISL FSPL and noise.} We analyze a fixed number of $19$ cooperative satellites and calculate the spectral efficiency by designating different satellites as the MN satellite. \textcolor{black}{Satellites are initially indexed in ascending order based on their distance from the user. Once the MN satellite is selected, the remaining satellites are re-indexed in ascending order based on their distance to the MN satellite. This ensures that the MN satellite collaborates with the $M-1$ nearest satellites.}
As discussed in Section \ref{MN Satellite Selection}, the MN satellite offers two key advantages: (1) it possesses at least one full instantaneous CSI (it's own), and (2) it is unaffected by ISL FSPL and noise. Consequently, the signal received by the MN satellite is more reliable, making it preferable to select the satellite with the strongest uplink channel.

\textcolor{black}{Additionally, the MN satellite should also be the one closest to the other satellites. This observation is supported by the results shown in Fig. \ref{fig:MN selection}. At lower ISL frequencies (1\,\text{THz}), where ISL noise is negligible, the spectral efficiency remains nearly the same across all MN satellite indices. However, as the ISL frequency increases to higher levels (15\,\text{THz}), ISL noise becomes more significant. In all scenarios, the four nearest satellites achieve the highest spectral efficiency due to their more reliable channels. Minor variations in spectral efficiency among these satellites arise from differences in the MN satellite’s distance to other satellites. For example, with perfect CSI at $10\,\text{THz}$, the fourth satellite slightly outperforms the first three because it has a shorter ISL distance to the other satellites. Furthermore, as the ISL frequency increases to $30\,\text{THz}$, the impact of ISL noise becomes even more pronounced. This leads to a sharper decline in the spectral efficiency of the fifth satellite and those farther away, as they have larger user-satellite distances.}

\begin{figure}[t]
    \centering
    \includegraphics[scale=0.34, trim={0cm 0cm 0cm 1.2cm}, clip]{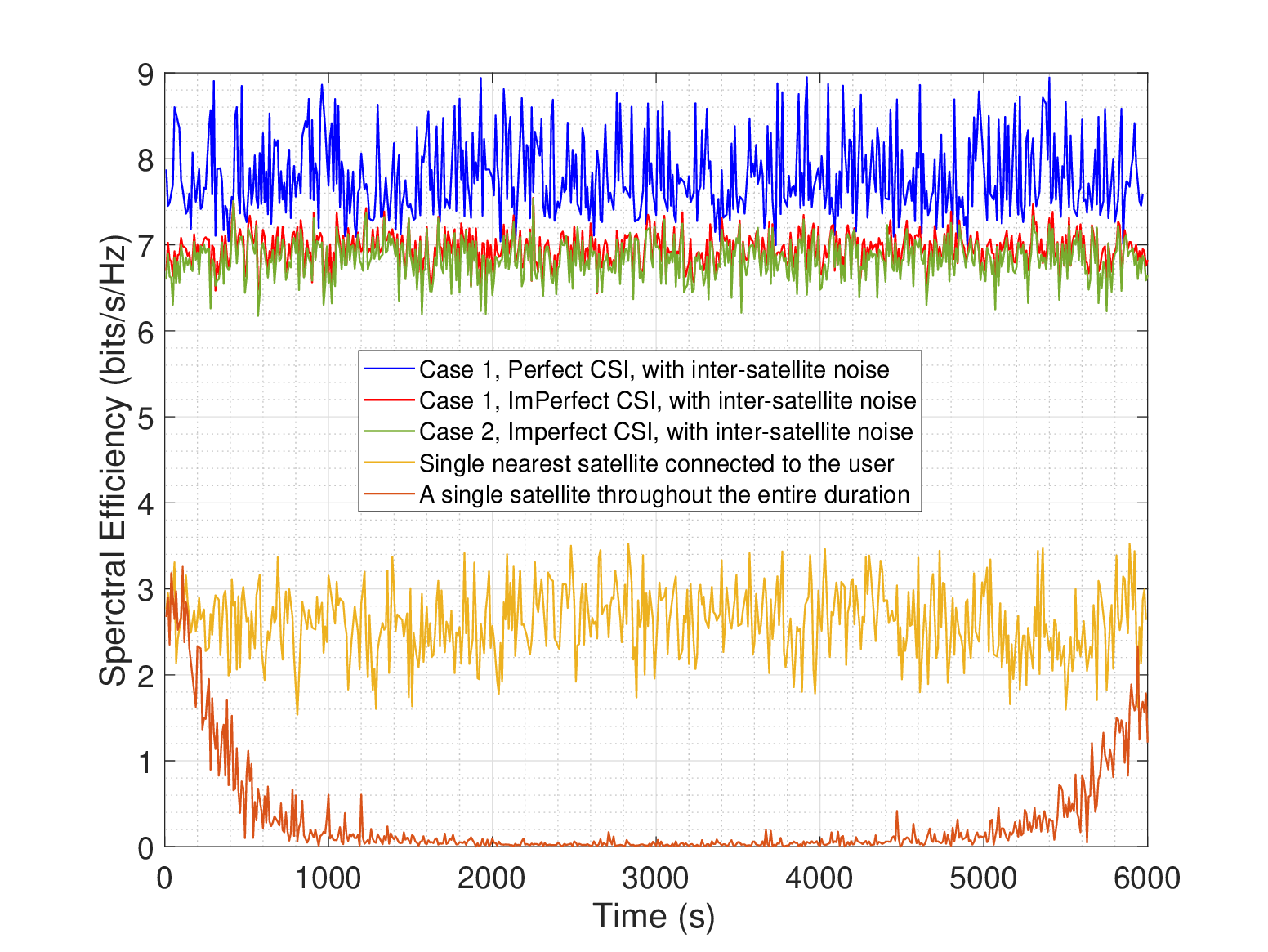}
    \caption{Spectral efficiency over a 100-minute duration.}
    \label{fig:1}
\end{figure}

\textcolor{black}{Fig. \ref{fig:1} illustrates the system's spectral efficiency for 19 satellites over a 6000-second duration of satellite movement for both cases of instantaneous CSI and statistical CSI sharing, referred to as Case 1 and Case 2, respectively. It is evident that multi-satellite MIMO systems can significantly enhance the average spectral efficiency, increasing it from $2.8\,\text{bits/s/Hz}$ to $8\,\text{bits/s/Hz}$.}

\textcolor{black}{In a single-satellite scenario, when the signal drops below a certain threshold, the connection is handed over to the next visible LEO satellite. Since the service time depends on the visibility of a single satellite, it typically lasts only a few minutes in LEO systems. In contrast, with multi-satellite MIMO, the user connects to a cluster of satellites, extending the service time to the visibility duration of the entire cluster, significantly reducing the handover rate \cite{abdelsadek2022distributed}.
Fig. ~\ref{fig:1} demonstrates that even with frequent handovers at each time step, the performance of single-satellite systems remains far below that of MIMO systems. Additionally, this approach requires nearly 50 handovers during a satellite's visibility period.
On the other hand, if handovers are not considered and only the nearest satellite to the user is selected for the entire satellite duration, the spectral efficiency peaks for about 2 minutes while the satellite remains within the visibility region for nearly 17 minutes. This indicates that, in non-cooperative systems, handovers would be required every 2 minutes to maintain the capacity of the nearest satellite. Therefore, satellite cooperation can greatly enhance system performance compared to traditional SatCom approaches.
Furthermore, as expected, Case 2, with statistical CSI sharing within a cluster, exhibits slightly lower capacity compared to Case 1, with instantaneous CSI sharing within a cluster.}

\begin{figure}[t]
    \centering
    \includegraphics[scale=0.34]{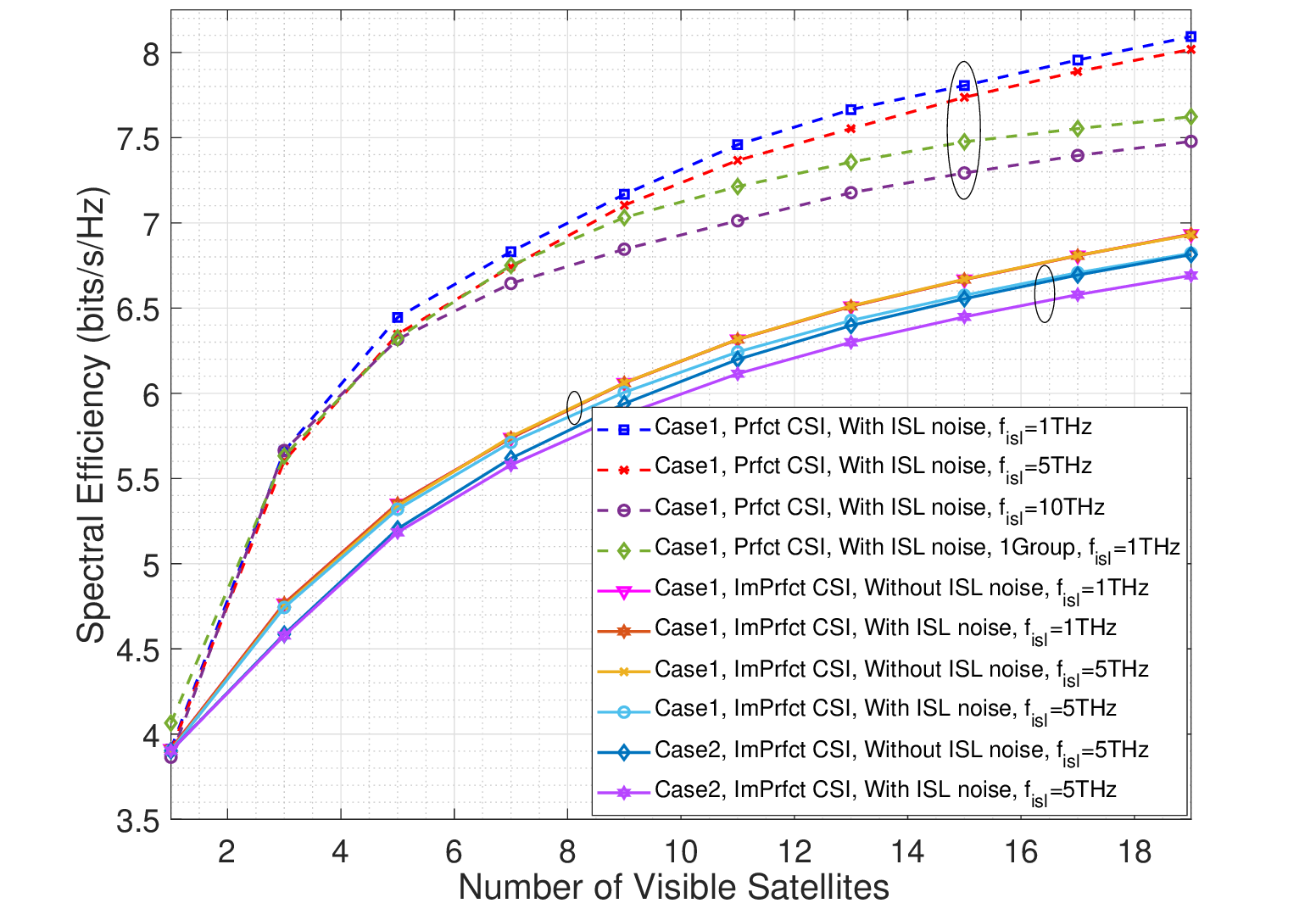}
    \caption{Spectral efficiency vs. number of satellites in a cluster.}
    \label{fig:2}
\end{figure}

\textcolor{black}{Fig. \ref{fig:2} illustrates the average spectral efficiency of the system as a function of the number of visible satellites for different ISL carrier frequencies in both cases. Generally, as the number of satellites increases, the diversity gain improves, leading to higher spectral efficiency.}
However, the spectral efficiency does not increase linearly due to the selection of satellites in a cluster based on their proximity to the user. Incorporating distant satellites may not provide significant performance improvements and could introduce high overhead.
For instance, assuming perfect CSI, Case 1 achieves a spectral efficiency of $7.2\,\text{bits/s/Hz}$ with $M=9$ satellites. Doubling the number of satellites to $M=18$ results in a spectral efficiency increase to $8\,\text{bits/s/Hz}$, representing only a $10\%$ improvement despite the doubled satellite count.
Thus, selecting an optimal number of satellites in a cluster is critical for maximizing system spectral efficiency while minimizing overhead. Optimizing constellation parameters is essential to balance spectral efficiency and cost, making this an intriguing area for future research.

\textcolor{black}{The first four dashed-line plots in Fig. \ref{fig:2} represent Case 1, which assumes perfect CSI for various ISL frequencies. It can be observed that as the ISL frequency increases, the spectral efficiency decreases. This occurs because higher ISL frequencies lead to increased ISL noise due to the larger ISL bandwidth.
Furthermore, this figure shows that selecting satellites from two groups rather than one results in higher capacity. This is because having two groups increases satellite density, enabling the selection of better-positioned satellites relative to the user.
In the solid-line plots, the spectral efficiency with and without considering ISL noise exhibits minimal differences, as ISL noise at these frequencies is negligible. However, as the ISL frequency increases from $1\,\text{THz}$ to $5\,\text{THz}$, ISL noise becomes significant, leading to reduced system performance when ISL noise is taken into account.
Moreover, this figure confirms the results shown in Fig. \ref{fig:1}, demonstrating that Case 1 outperforms Case 2 across all satellite numbers due to its access to more detailed channel information. In contrast, Case 2 relies solely on long-term channel information.}

\begin{figure}[t]
    \centering
    \includegraphics[scale=0.34]{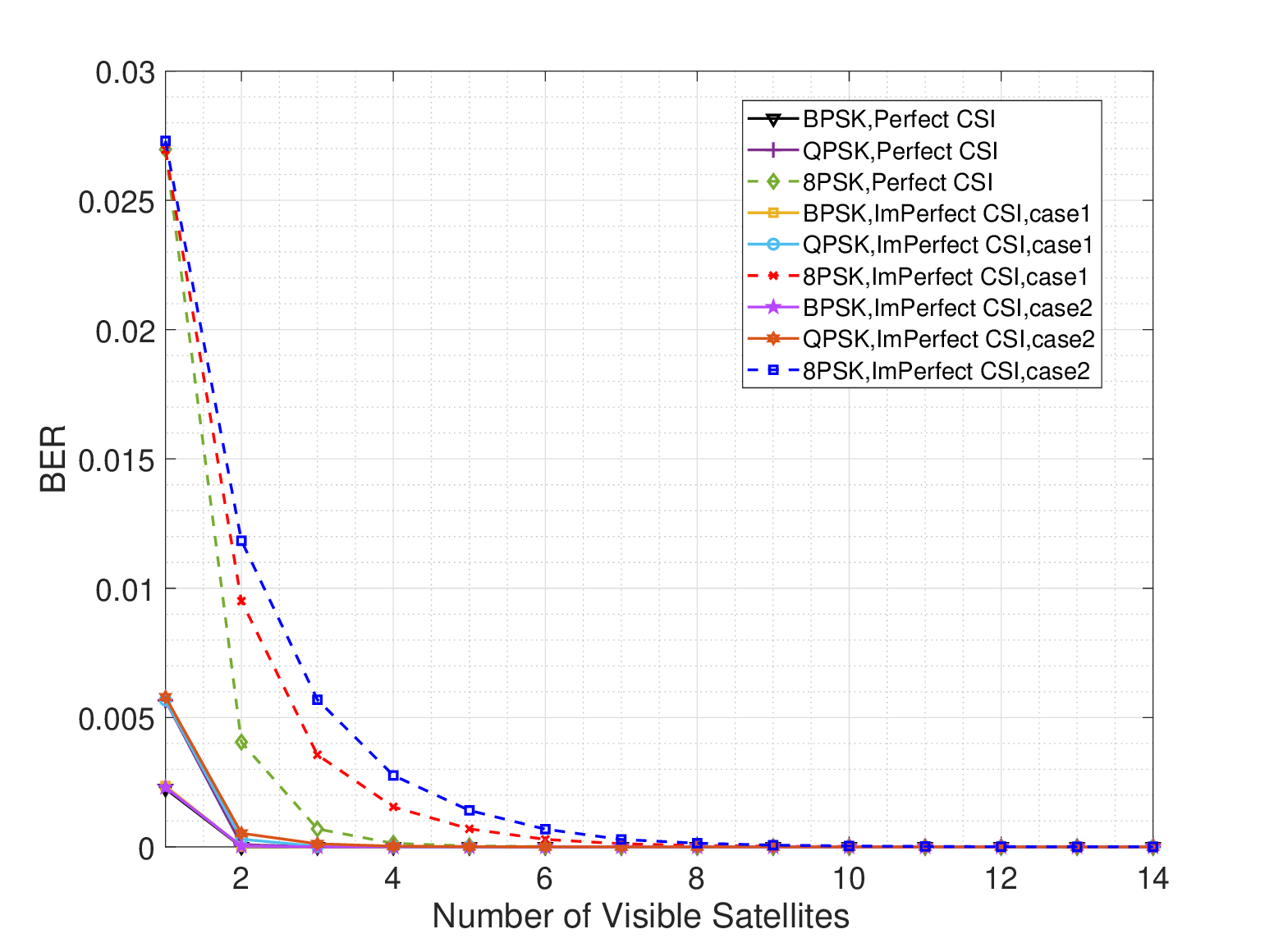}
    \caption{BER performance vs. number of satellites in a cluster.}
    \label{fig:3}
\end{figure}
\textcolor{black}{Fig. \ref{fig:3} illustrates the BER performance of the system as a function of the number of visible satellites in a cluster, considering both Perfect CSI and Imperfect CSI scenarios for different modulation schemes (BPSK, QPSK, and 8PSK). As the number of visible satellites increases, the BER decreases across all cases. This improvement is attributed to the increased diversity gain, as having more satellites in the cluster enhances signal reliability and robustness against noise.
In the case of Imperfect CSI, the BER is higher compared to Perfect CSI due to the lack of full channel information, which results in suboptimal signal detection and processing. Furthermore, Case 1 demonstrates better BER performance compared to Case 2, as it benefits from access to more channel information. This is particularly evident for higher-order modulation schemes like 8PSK.
For all scenarios, the BER converges to near zero as the number of visible satellites exceeds 9. This implies that with more than 9 satellites, it becomes feasible to use 8PSK modulation at each time step without errors. This result highlights that the diversity gain provided by additional satellites is sufficient to overcome channel impairments, even under Imperfect CSI conditions.}

\begin{figure}[t]
    \centering
   \includegraphics[scale=0.34]{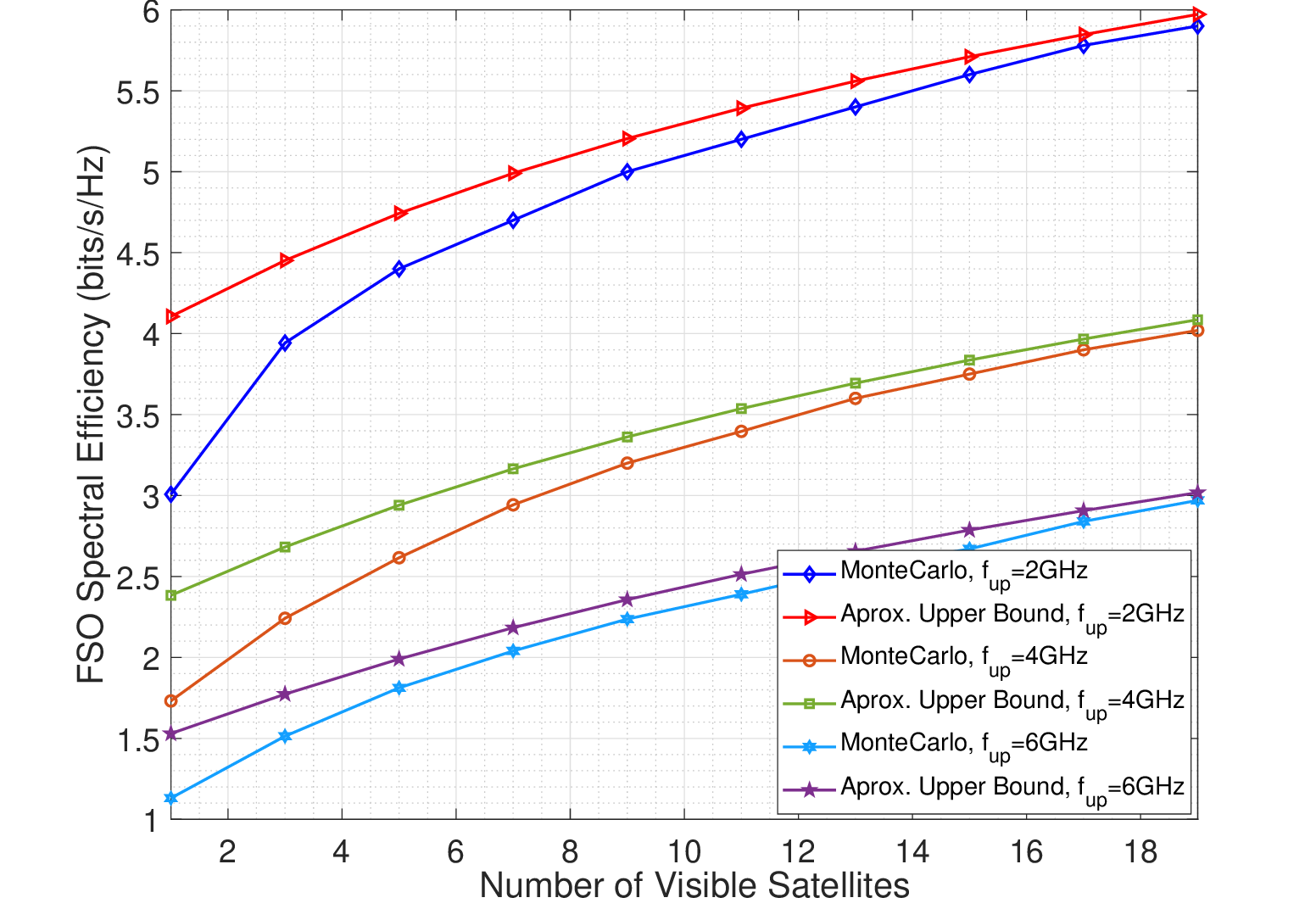}
   \caption{Validation of the approximation of FSO ergodic capacity upper bound.}
    \label{fso validation}
\end{figure}

\begin{figure}[t]
    \centering
   \includegraphics[scale=0.34]{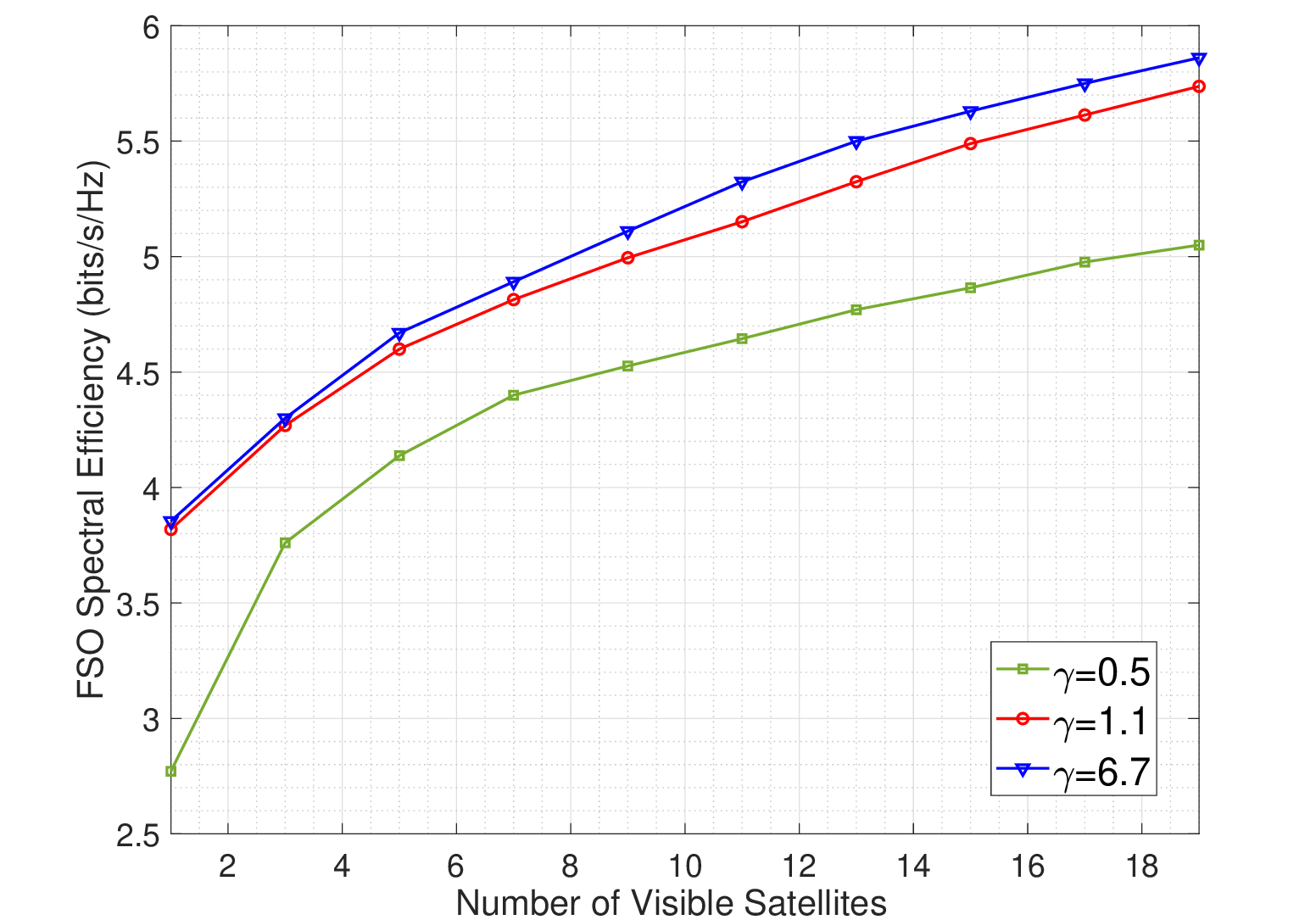}
    \caption{Spectral efficiency vs number of satellites for different values of $\gamma$.}
    \label{fig_z_1}
\end{figure}

The verification of our FSO upper bound approximation, in (\ref{eq42}), is depicted in Fig. \ref{fso validation}. In this simulation, we assume that all satellites are approximately at the same distance from the user, leading to equal channel variance. Consequently, the covariance matrix is diagonal. Given the average user-satellite distance of $550\,\text{km}$, the uplink FSPL is calculated as $\beta_{\text{up}} = 2.12 \times 10^{15}$. Thus, we model a Rician channel with a mean vector of $\frac{1}{\sqrt{\beta_{\text{up}}}}(\alpha_r + j\alpha_t) \mb{u}$ and a covariance matrix of $\frac{1}{\beta_{\text{up}}} \mb{I}$. Moreover, we consider an average ISL distance of $900\,\text{km}$. So, we have  $\beta_{\text{isl}} = 5.29 \times 10^{25}$.   
Fig. \ref{fso validation} compares Monte Carlo simulations with our approximated upper bound from (\ref{eq42}) for three different uplink carrier frequencies, demonstrating that (\ref{eq42}) closely approximates the upper bound for the FSO spectral efficiency in our scenario.
\textcolor{black}{Fig. \ref{fig_z_1} illustrates the spectral efficiency as a function of the number of visible satellites, considering different levels of pointing error. It is evident that the spectral efficiency improves as the impact of the pointing error decreases (i.e., for higher values of $\gamma$).}


\section{conclusion}\label{conclusion}
This paper investigated a practical scenario based on 3GPP standards for uplink transmission from a single-antenna mobile phone to a cluster of satellites that cooperate to detect the received signal. The cooperation was enabled through ISLs, utilizing either THz or FSO communication technologies. The study first identified the optimal satellite to act as the MN satellite, which was responsible for signal processing.
For THz communications, two scenarios were considered: instantaneous CSI sharing and statistical CSI sharing among the satellites. In both cases, a closed-form approximation for the ergodic capacity was proposed. For FSO communications, the pointing error loss was modeled, and a closed-form approximate upper bound for the ergodic capacity was derived. Furthermore, the impact of pointing errors on the FSO ISL capacity was analyzed.
Numerical results revealed that the cooperative satellite design significantly outperforms traditional SatCom schemes, where only a single satellite is connected to the user. The results also emphasized that system performance depends not only on the number of satellites in the cluster but also on constellation parameters, such as satellite density.

\appendices
\section{}\label{FirstAppendix}
$\mb{Lemma \ 1}$: Assuming $\mb{x}$ is an $M \times 1$ complex vector, the following inequality holds:
\begin{equation}
    _0F_1(M, \mb{x}^H\mb{x}) \leq \ _0F_0\left(\frac{\mb{x}^H\mb{x}}{M}\right).
\end{equation}
Equality is achieved as $M \rightarrow \infty$ \cite{ratnarajah2004spatially}.\\
By setting $\mb{x} = \sqrt{\frac{\Omega}{\sigma_h^2}}\mb{h}$, we obtain,
\begin{equation}
    _0F_1\left(M, \frac{\Omega w}{\sigma_h^2}\right) \leq \ _0F_0\left(\frac{\Omega w}{M\sigma_h^2}\right).
\end{equation}
The hypergeometric function can now be rewritten as $_0F_0\left(\frac{\Omega w}{M\sigma_h^2}\right) = e^{\frac{\Omega w}{M\sigma_h^2}}$ \cite{james1964distributions}, providing the following upper bound for the ergodic capacity:
\begin{equation}\label{upper bound 1}
    C_e \leq \frac{e^{-\Omega}}{(M-1)!(\sigma_h^2)^{M}} \int_0^{\infty} \log_2(\rho w + 1) e^{w\left(\frac{\Omega - M}{M\sigma_h^2}\right)} w^{M-1} \, \text{d}w.
\end{equation}
There is no closed-form solution for the integral in (\ref{upper bound 1}) using elementary functions. However, numerical integration methods, such as Gauss-Legendre quadrature, can be employed to obtain an accurate approximation for a wide range of functions. Alternatively, focusing on the low-SNR regime, which is relevant in SatComs due to significant path loss, offers another approach. In this regime, since $\rho w \ll 1$, we can use the approximation $\log_2(\rho w + 1) \approx \rho w$. By substituting $\log_2(\rho w + 1)$ with $\rho w$, the integral simplifies to $\rho \int_0^{\infty} e^{-w(\frac{M - \Omega}{M\sigma_h^2})} w^M \, \text{d}w$, providing an approximation of the original expression. This simplified integral has a closed-form solution, as detailed in \cite{zwillinger2007table} (section 3.381):
\begin{equation}\label{integral}
    \int_0^{\infty} e^{-w(\frac{M - \Omega}{M\sigma_h^2})} w^M \, \text{d}w = \frac{(M\sigma_h^2)^{M+1} M!}{(M - \Omega)^{M+1}}.
\end{equation}
This results in an approximation for the ergodic capacity given by,
\begin{align}
    C_e \approx \frac{\rho \sigma_h^2 e^{-\Omega} M^{M+2}}{(M - \Omega)^{M+1}}.
\end{align}

\normalem
\bibliographystyle{IEEEtran}
\bibliography{Ref1}

\begin{thebibliography}{10}
\providecommand{\url}[1]{#1}
\csname url@samestyle\endcsname
\providecommand{\newblock}{\relax}
\providecommand{\bibinfo}[2]{#2}
\providecommand{\BIBentrySTDinterwordspacing}{\spaceskip=0pt\relax}
\providecommand{\BIBentryALTinterwordstretchfactor}{4}
\providecommand{\BIBentryALTinterwordspacing}{\spaceskip=\fontdimen2\font plus
\BIBentryALTinterwordstretchfactor\fontdimen3\font minus \fontdimen4\font\relax}
\providecommand{\BIBforeignlanguage}[2]{{%
\expandafter\ifx\csname l@#1\endcsname\relax
\typeout{** WARNING: IEEEtran.bst: No hyphenation pattern has been}%
\typeout{** loaded for the language `#1'. Using the pattern for}%
\typeout{** the default language instead.}%
\else
\language=\csname l@#1\endcsname
\fi
#2}}
\providecommand{\BIBdecl}{\relax}
\BIBdecl

\bibitem{10179219}
J.~Heo, S.~Sung, H.~Lee, I.~Hwang, and D.~Hong, ``{MIMO Satellite Communication Systems: A Survey From the PHY Layer Perspective},'' \emph{IEEE Communications Surveys \& Tutorials}, vol.~25, no.~3, pp. 1543--1570, 2023.

\bibitem{9852737}
H.~Al-Hraishawi, H.~Chougrani, S.~Kisseleff, E.~Lagunas, and S.~Chatzinotas, ``{A Survey on Nongeostationary Satellite Systems: The Communication Perspective},'' \emph{IEEE Communications Surveys \& Tutorials}, vol.~25, no.~1, pp. 101--132, 2023.

\bibitem{10646360}
Z.~M. Bakhsh, Y.~Omid, G.~Chen, F.~Kayhan, Y.~Ma, and R.~Tafazolli, ``{Multi-Satellite MIMO Systems for Direct Satellite-to-Device Communications: A Survey},'' \emph{IEEE Communications Surveys \& Tutorials}, pp. 1--1, 2024.

\bibitem{9351765}
Q.~Chen, G.~Giambene, L.~Yang, C.~Fan, and X.~Chen, ``{Analysis of Inter-Satellite Link Paths for LEO Mega-Constellation Networks},'' \emph{IEEE Transactions on Vehicular Technology}, vol.~70, no.~3, pp. 2743--2755, 2021.

\bibitem{chen2021modeling}
Q.~Chen, L.~Yang, X.~Liu, B.~Cheng, J.~Guo, and X.~Li, ``Modeling and analysis of inter-satellite link in {LEO} satellite networks,'' in \emph{2021 13th International Conference on Communication Software and Networks (ICCSN)}, Chongqing, China, 2021, pp. 134--138.

\bibitem{maamar2016study}
B.~Maamar and X.~Mai, ``Study and analysis of optical intersatellite links,'' \emph{International Journal of Electronics and Communication Engineering}, vol.~10, no.~6, pp. 717--721, 2016.

\bibitem{nie2021channel}
S.~Nie and I.~F. Akyildiz, ``Channel modeling and analysis of inter-small-satellite links in terahertz band space networks,'' \emph{IEEE Transactions on Communications}, vol.~69, no.~12, pp. 8585--8599, 2021.

\bibitem{ding2016analysis}
Y.~Ding, S.~Gao, X.~Shi, and H.~Wu, ``Analysis of inter-satellite {T}erahertz communication link,'' in \emph{3rd International Conference on Wireless Communication and Sensor Networks (WCSN 2016)}.\hskip 1em plus 0.5em minus 0.4em\relax Atlantis Press, 2016, pp. 180--183.

\bibitem{Salem2024}
H.~B. Salem and Others, ``{Uplink Soft Handover for LEO Constellations: How Strong the Inter-Satellite Link Should Be},'' \emph{arXiv preprint}, vol. arXiv:2403.15131, 2024, available at: \url{https://arxiv.org/abs/2403.15131}.

\bibitem{10197164}
N.~Ye, X.~Cao, X.~Ding, J.~Li, D.~Zhao, and Q.~Ouyang, ``{Multi-Connection to the Sky: Energy-Efficient Beamforming for Multi-Satellite Uplink Transmission With Lens Antenna Array},'' \emph{IEEE Transactions on Green Communications and Networking}, vol.~7, no.~4, pp. 1836--1848, 2023.

\bibitem{10279447}
M.~Röper, B.~Matthiesen, D.~Wübben, P.~Popovski, and A.~Dekorsy, ``{Robust Precoding via Characteristic Functions for VSAT to Multi-Satellite Uplink Transmission},'' in \emph{ICC 2023 - IEEE International Conference on Communications}, 2023, pp. 6281--6286.

\bibitem{chen2024direct}
E.~Chen, R.-A. Pitaval, B.~M. Popovi\'{c}, and Y.~Qin, ``{Direct Satellite Access Using Multi-Dimensional Constellations},'' in \emph{PIMRC 2024 - IEEE International Symposium on Personal, Indoor and Mobile Radio Communications}, 2024.

\bibitem{3GPPrel15}
``{3rd generation partnership project; technical specification group radio access network; study on new radio (NR) to support non terrestrial networks (release 15)},'' \emph{Tech. Rep. {3GPP} TR 38.811 V15.0.0}, Jun. 2018.

\bibitem{roper2022beamspace}
M.~Röper, B.~Matthiesen, D.~Wübben, P.~Popovski, and A.~Dekorsy, ``{Beamspace MIMO for Satellite Swarms},'' in \emph{2022 IEEE Wireless Communications and Networking Conference (WCNC)}, 2022, pp. 1307--1312.

\bibitem{9814655}
Y.~Wang, Z.~Zheng, M.~Zeng, and Z.~Fei, ``{Coordinated precoding for Multi-Satellite Communications: A Deterministic Equivalent Approach},'' in \emph{2022 IEEE International Conference on Communications Workshops (ICC Workshops)}, 2022, pp. 1165--1170.

\bibitem{li2022performance}
Z.~Li, S.~Chen, and S.~Han, ``{Performance Analysis of Downlink Multisatellite Joint Service System Toward SAGOI-Net},'' \emph{IEEE Internet of Things Journal}, vol.~10, no.~11, pp. 9366--9374, 2023.

\bibitem{abdelsadek2022distributed}
M.~Y. Abdelsadek, G.~K. Kurt, and H.~Yanikomeroglu, ``{Distributed Massive MIMO for LEO Satellite Networks},'' \emph{IEEE Open Journal of the Communications Society}, vol.~3, pp. 2162--2177, 2022.

\bibitem{10061620}
M.~Y. Abdelsadek, G.~Karabulut-Kurt, H.~Yanikomeroglu, P.~Hu, G.~Lamontagne, and K.~Ahmed, ``{Broadband Connectivity for Handheld Devices via LEO Satellites: Is Distributed Massive MIMO the Answer?}'' \emph{IEEE Open Journal of the Communications Society}, vol.~4, pp. 713--726, 2023.

\bibitem{omid2023oncapacity}
Y.~Omid, Z.~Mashayekh~Bakhsh, F.~Kayhan, Y.~Ma, F.~Wang, and R.~Tafazolli, ``{On Capacity of Handheld to Multi-Satellite Communication},'' in \emph{PIMRC 2023: IEEE International Symposium on Personal, Indoor and Mobile Radio Communications}, July 2023, {Accepted}.

\bibitem{maaref2007joint}
A.~Maaref and S.~Aissa, ``{Joint and marginal eigenvalue distributions of (non) central complex Wishart matrices and PDF-based approach for characterizing the capacity statistics of MIMO Ricean and Rayleigh fading channels},'' \emph{IEEE Transactions on Wireless Communications}, vol.~6, no.~10, pp. 3607--3619, 2007.

\bibitem{ratnarajah2004spatially}
T.~Ratnarajah, ``{Spatially correlated MIMO Rician channel capacity},'' in \emph{Conference Record of the Thirty-Eighth Asilomar Conference on Signals, Systems and Computers, 2004.}, vol.~1.\hskip 1em plus 0.5em minus 0.4em\relax IEEE, 2004, pp. 1188--1192.

\bibitem{omid2023spacemimo}
Y.~Omid, Z.~Mashayekh~Bakhsh, F.~Kayhan, Y.~Ma, and R.~Tafazolli, ``{Space MIMO: Direct Unmodified Handheld to Multi-Satellite Communication},'' in \emph{GLOBECOM 2023: IEEE Global Communications Conference}, August 2023, {Accepted}.

\bibitem{10496842}
Z.~An, Y.~Xu, A.~Tahir, J.~Wang, B.~Ma, G.~F. Pedersen, and M.~Shen, ``{Collaborative Learning-Based Modulation Recognition for 6G Multibeam Satellite Communication Systems via Blind and Semiblind Channel Equalization},'' \emph{IEEE Transactions on Aerospace and Electronic Systems}, vol.~60, no.~4, pp. 5226--5246, 2024.

\bibitem{10077720}
X.~Zhang, X.~Yue, T.~Li, Z.~Han, Y.~Wang, Y.~Ding, and R.~Liu, ``{A Unified NOMA Framework in Beam-Hopping Satellite Communication Systems},'' \emph{IEEE Transactions on Aerospace and Electronic Systems}, vol.~59, no.~5, pp. 5390--5404, 2023.

\bibitem{humadi2024distributed}
K.~Humadi, G.~K. Kurt, and H.~Yanikomeroglu, ``{Distributed Massive MIMO System with Dynamic Clustering in LEO Satellite Networks},'' \emph{arXiv preprint arXiv:2404.06024}, 2024.

\bibitem{maharjan2022atmospheric}
N.~Maharjan, N.~Devkota, and B.~W. Kim, ``Atmospheric effects on satellite--ground free space uplink and downlink optical transmissions,'' \emph{Applied Sciences}, vol.~12, no.~21, p. 10944, 2022.

\bibitem{10515772}
M.~Kumari, N.~Sharma, R.~Chauhan, K.~Joshi, and A.~Kumar, ``{Inter-satellite OWC Architecture for LEO-GEO Scenarios},'' in \emph{2024 International Conference on Distributed Computing and Optimization Techniques (ICDCOT)}, 2024, pp. 1--5.

\bibitem{duman2008coding}
T.~M. Duman and A.~Ghrayeb, \emph{Coding for MIMO communication systems}.\hskip 1em plus 0.5em minus 0.4em\relax John Wiley \& Sons, 2008.

\bibitem{ratnarajah2003topics}
T.~Ratnarajah, ``Topics in complex random matrices and information theory,'' Ph.D. dissertation, University of Ottawa (Canada), 2003.

\bibitem{james1964distributions}
A.~T. James, ``Distributions of matrix variates and latent roots derived from normal samples,'' \emph{The Annals of Mathematical Statistics}, vol.~35, no.~2, pp. 475--501, 1964.

\bibitem{9108615}
E.~Zedini, A.~Kammoun, and M.-S. Alouini, ``{Performance of Multibeam Very High Throughput Satellite Systems Based on FSO Feeder Links With HPA Nonlinearity},'' \emph{IEEE Transactions on Wireless Communications}, vol.~19, no.~9, pp. 5908--5923, 2020.

\bibitem{7553489}
H.~Kaushal and G.~Kaddoum, ``{Optical Communication in Space: Challenges and Mitigation Techniques},'' \emph{IEEE Communications Surveys \& Tutorials}, vol.~19, no.~1, pp. 57--96, 2017.

\bibitem{9822386}
T.~V. Nguyen, H.~D. Le, and A.~T. Pham, ``{On the Design of RIS–UAV Relay-Assisted Hybrid FSO/RF Satellite–Aerial–Ground Integrated Network},'' \emph{IEEE Transactions on Aerospace and Electronic Systems}, vol.~59, no.~2, pp. 757--771, 2023.

\bibitem{aladeloba2013optically}
A.~O. Aladeloba, ``Optically amplified free-space optical communication systems,'' Ph.D. dissertation, University of Nottingham, 2013.

\bibitem{9535285}
P.~K. Singya and M.-S. Alouini, ``{Performance of UAV-Assisted Multiuser Terrestrial-Satellite Communication System Over Mixed FSO/RF Channels},'' \emph{IEEE Transactions on Aerospace and Electronic Systems}, vol.~58, no.~2, pp. 781--796, 2022.

\bibitem{5452205}
S.~Jin, M.~R. McKay, C.~Zhong, and K.-K. Wong, ``{Ergodic Capacity Analysis of Amplify-and-Forward MIMO Dual-Hop Systems},'' \emph{IEEE Transactions on Information Theory}, vol.~56, no.~5, pp. 2204--2224, 2010.

\bibitem{mcshane1937jensen}
E.~J. McShane, ``{Jensen's inequality},'' 1937.

\bibitem{liang2021phasing}
J.~Liang, A.~U. Chaudhry, and H.~Yanikomeroglu, ``Phasing parameter analysis for satellite collision avoidance in starlink and kuiper constellations,'' in \emph{2021 IEEE 4th 5G World Forum (5GWF)}.\hskip 1em plus 0.5em minus 0.4em\relax IEEE, 2021, pp. 493--498.

\bibitem{omid2024reinforcement}
Y.~Omid, M.~Aristodemou, S.~Lambotharan, M.~Derakhshani, and L.~Hanzo, ``{Reinforcement Learning-Based Downlink Transmit Precoding for Mitigating the Impact of Delayed CSI in Satellite Systems},'' \emph{arXiv preprint}, 2024.

\bibitem{alkholidi2014fso}
\BIBentryALTinterwordspacing
A.~G. Alkholidi and K.~S. Altowij, ``Free space optical communications -- theory and practices,'' in \emph{Contemporary Issues in Wireless Communications}, M.~Khatib, Ed.\hskip 1em plus 0.5em minus 0.4em\relax IntechOpen, 2014, ch. unknown, p. unknown. [Online]. Available: \url{https://www.intechopen.com/chapters/47585}
\BIBentrySTDinterwordspacing

\bibitem{zwillinger2007table}
D.~Zwillinger and A.~Jeffrey, \emph{Table of integrals, series, and products}.\hskip 1em plus 0.5em minus 0.4em\relax Elsevier, 2007.

\end{thebibliography}
\end{document}